\documentclass[reqno,11pt]{amsart}
\usepackage[utf8]{inputenc}
\usepackage{amscd}
\usepackage{slashed}
\usepackage{amssymb}
\usepackage{esint}
\usepackage{enumitem}
\usepackage[off]{auto-pst-pdf} 
\usepackage{epsfig}
\usepackage{pst-grad} 
\usepackage{pst-plot} 
\usepackage[mathscr]{eucal}
\numberwithin{equation}{section}
\allowdisplaybreaks[1]

\newcommand{\SetFigFont}[3]{}

\title[Noether-Like Theorems for Causal Variational Principles]{Noether-Like Theorems for \\
Causal Variational Principles}
\author[F.\ Finster]{Felix Finster}
\author[J.\ Kleiner]{Johannes Kleiner \\ \\ June 2015}
\address{Fakult\"at f\"ur Mathematik \\ Universit\"at Regensburg \\ D-93040 Regensburg \\ Germany}
\email{finster@ur.de, Johannes.Kleiner@ur.de}

\newtheorem{Def}{Definition}[section]
\newtheorem{Thm}[Def]{Theorem}
\newtheorem{Prp}[Def]{Proposition}
\newtheorem{Lemma}[Def]{Lemma}
\newtheorem{Remark}[Def]{Remark}
\newtheorem{Corollary}[Def]{Corollary}

\newcommand{\Thanks}{\vspace*{.5em} \noindent \thanks}
\newcommand{\beq}{\begin{equation}}
\newcommand{\eeq}{\end{equation}}
\newcommand{\Proof}{\begin{proof}}
\newcommand{\QED}{\end{proof} \noindent}
\newcommand{\QEDrem}{\ \hfill $\Diamond$}

\newcommand{\la}{\langle}
\newcommand{\ra}{\rangle}

\newcommand{\Sl}{\mbox{$\prec \!\!$ \nolinebreak}}
\newcommand{\Sr}{\mbox{\nolinebreak $\succ$}}

\newcommand{\C}{\mathbb{C}}
\newcommand{\R}{\mathbb{R}}
\newcommand{\1}{\mbox{\rm 1 \hspace{-1.05 em} 1}}

\newcommand{\N}{\mathbb{N}}

\newcommand{\F}{\mathscr{F}} 

\newcommand{\Pdd}{\mbox{$\partial$ \hspace{-1.2 em} $/$}}
\newcommand{\slsh}{\mbox{ \hspace{-1.13 em} $/$}}

\newcommand{\na}{n_{\mathrm{a}}}

\renewcommand{\H}{\mathscr{H}}
\newcommand{\scrM}{\mycal M}
\newcommand{\scrN}{\mycal N}
\newcommand{\Lin}{\text{\rm{L}}}

\DeclareFontFamily{OT1}{rsfso}{}
\DeclareFontShape{OT1}{rsfso}{m}{n}{ <-7> rsfso5 <7-10> rsfso7 <10-> rsfso10}{}
\DeclareMathAlphabet{\mycal}{OT1}{rsfso}{m}{n}

\DeclareMathOperator{\re}{Re}
\DeclareMathOperator{\im}{Im}
\DeclareMathOperator{\Tr}{Tr}
\DeclareMathOperator{\tr}{tr}

\DeclareMathOperator{\supp}{supp}

\renewcommand{\L}{{\mathcal{L}}}
\newcommand{\Sact}{{\mathcal{S}}}
\newcommand{\T}{{\mathcal{T}}}

\newcommand{\scrU}{{\mathscr{U}}}
\newcommand{\scrA}{{\mathscr{A}}}

\begin{document}
\maketitle

\begin{abstract}
The connection between symmetries and conservation laws
as made by Noether's theorem is extended to the
context of causal variational principles and causal fermion systems.
Different notions of continuous symmetries are introduced.
It is proven that these symmetries give rise to corresponding conserved quantities,
expressed in terms of so-called surface layer integrals.
In a suitable limiting case, the Noether-like theorems for causal fermion systems
reproduce charge conservation and the
conservation of energy and momentum in Minkowski space.
Thus the conservation of charge and energy-momentum are found to be
special cases of general conservation laws which are intrinsic to causal fermion systems.
\end{abstract}

\tableofcontents

\section{Introduction and Statement of Results}
In modern physics, the connection between symmetries and conservation laws
is of central importance. For continuous symmetries, this connection is made mathematically precise
by Noether's theorem~\cite{noetheroriginal}.
In recent years, the theory of causal fermion systems was proposed as an approach
to describe fundamental physics. Giving quantum mechanics, general relativity and quantum field theory
as limiting cases, it is a candidate for a unified physical theory (see the review~\cite{dice2014}
and the references therein).
In the present article, we explore symmetries and the resulting conservation laws
in the framework of causal fermion systems.
We prove that there are indeed conservation laws, which however have a
structure which is quite different from that of the classical Noether theorem.
These conservation laws are so general that they apply to ``quantum space-times'' which cannot
be approximated by a Lorentzian manifold.  We prove that in the proper limiting case,
our conservation laws simplify to charge conservation and the conservation of energy and momentum
in Minkowski space.
Thus the conservation laws of charge and energy-momentum can be viewed as
special cases of more general conservation laws which are intrinsic to causal fermion systems.

In order to make the paper easily accessible and self-contained, we develop our concepts step by step.
Section~\ref{secprelim} provides the necessary background:
After a brief review of the classical Noether theorem (Section~\ref{secclass}), we
introduce causal variational principles in the compact setting (a mathematical simplification
of the setting of causal fermion systems; see Section~\ref{secintroc}).
This makes it possible to describe the mathematical structure of
our conservation laws in the simplest possible situation (Section~\ref{secsli}).
The central point is that, instead of surface integrals, we work with integrals
over ``thin layers of finite thickness,'' referred to as {\em{surface layer integrals}}
(see Figure~\ref{fignoether1} on page~\pageref{fignoether1}).

After these preparations, in Section~\ref{seccompact} we prove conservation laws
for causal variational principles in the compact setting.
We distinguish two different kinds of symmetries: symmetries of the
Lagrangian (see Definition~\ref{defsymmrho} and Theorem~\ref{thmsymmum})
and symmetries of the universal measure (see
Definition~\ref{defsymmlagr} and Theorem~\ref{thmsymmlag}).
These symmetries and the corresponding conservation laws can be combined
in so-called {\em{generalized integrated symmetries}} (see Definition~\ref{defgis}
and Theorem~\ref{thmsymmgis}).

In Section~\ref{seccfs} we generalize the previous results to the setting
of causal fermion systems. After a brief introduction to the mathematical setup
(Section~\ref{seccfsbasic}), we derive corresponding Noether-like theorems
(see Theorem~\ref{thmsymmgis2}, Corollary~\ref{corsymmlag} and Corollary~\ref{corsymmum}
in Section~\ref{seccfsnoether}).
In the following Sections~\ref{secexcurrent} and~\ref{secexEM}, we work out 
examples which give the correspondence to current conservation (Theorem~\ref{thmcurrentmink})
and to the conservation of energy-momentum (Corollary~\ref{corEMcons}).
In Section~\ref{secremark}, the mathematical assumptions and the physical picture is discussed and
clarified by a few remarks. 
In Section~\ref{secexrho} it is explained why the conservation laws corresponding to
symmetries of the universal measure are trivially satisfied in Minkowski space and do not capture any
interesting dynamical information.
Finally, in Section~\ref{secoutlook} we give an
outlook on potential implications for the collapse of the quantum mechanical wave functions
(as proposed in~\cite[Section~3]{dice2010} and~\cite[Section~7]{dice2014})
and for the mechanism of microscopic mixing of the wave functions (as introduced in~\cite[Section~3]{qft}).

\section{Preliminaries} \label{secprelim}
\subsection{The Classical Noether Theorem} \label{secclass}
We now briefly review Noether's theorem~\cite{noetheroriginal} in the form most suitable for our
purposes (similar formulations are found in~\cite[Section~13.7]{goldstein} or~\cite[Chapter~III]{barutbook}).
For simplicity, we begin in four-dimensional Minkowski space~$\scrM$.
In the Lagrangian formulation of classical field theory, one seeks for 
critical points of an action of the form
\[ \Sact = \int_{\scrM} \L \big( \psi(x), \psi_{,j}(x), x\big)\: d^4x \]
(where~$\psi$ is for example a scalar, tensor or spinor field, and~$\psi_{,j} \equiv \partial_j \psi$
denotes the partial derivative).
The critical field configurations satisfy the Euler-Lagrange (EL) equations
\beq \label{ELclass}
\frac{\partial \L}{\partial \psi} - \frac{\partial}{\partial x^j} \left( \frac{\partial \L}{\partial \psi_{,j}} \right) = 0 \:.
\eeq
Symmetries are formulated in terms of variations of the
field and the space-time coordinates. More precisely, for given~$\tau_{\max}>0$
we consider smooth families~$(\psi_\tau)$ and~$(x_\tau)$ parametrized
by~$\tau \in (-\tau_{\max}, \tau_{\max})$ with~$\psi_\tau|_{\tau=0}=\psi$ and~$x_\tau|_{\tau=0}=x$.
We assume that these variations describe a {\em{symmetry of the action}},
meaning that for every compact space-time region~$\Omega \subset \scrM$ and
every field configuration~$\psi$ the equation
\beq \label{symm}
\int_\Omega \L\big( \psi(x), \psi_{,j}(x), x\big)\: d^4x
= \int_{\Omega'} \L\big( \psi_\tau(y), (\psi_\tau)_{,j}(y), y \big)\: d^4y
\eeq
holds for all~$\tau \in (-\tau_{\max}, \tau_{\max})$,
where~$\Omega' = \{x_\tau \,|\, x \in \Omega\}$ is the transformed region.
The corresponding {\em{Noether current}}~$J$ is defined by
\[ J^k = \frac{\partial \L}{\partial \psi_{,k}} \,\delta \psi
+ \L\: \delta x^k - \frac{\partial \L}{\partial \psi_{,k}}\: \partial_j \psi\: \delta x^j \:, \]
where~$\delta x$ and~$\delta \psi$ are the first variations
\[ \delta x := \frac{d}{d\tau}\, x_\tau |_{\tau=0} \qquad \text{and} \qquad
\delta \psi(x) := \frac{d}{d\tau}\, \psi_\tau(x_\tau) |_{\tau=0} \:. \]
Noether's theorem states that if~$\psi$ satisfies the EL equations, then
the Noether current is divergence-free,
\[ \partial_k J^k = 0 \:. \]
Using the Gau{\ss} divergence theorem, one may integrate this equation to obtain
a corresponding conserved quantity. To this end, one chooses a space-time region~$\Omega$
whose boundary~$\partial \Omega$ consists of two space-like hypersurfaces~$\scrN_1$
and~$\scrN_2$. Then
\beq \label{conserve}
\int_{\scrN_1} J^k \nu_k\: d\mu_{\scrN_1}(x) = \int_{\scrN_2} J^k \nu_k\: d\mu_{\scrN_2}(x) \:,
\eeq
where~$\nu$ is the future-directed normal, and~$d\mu_{\scrN_{1\!/\!2}}$ is the induced volume measure
(if~$\Omega$ is unbounded, one needs to assume suitable decay of~$J^k$ at infinity).

We now mention two well-known applications of Noether's theorem which will be most relevant here.
The first application is to consider the Lagrangian of a quantum mechanical wave function~$\psi$
(like the Schr\"odinger, Klein-Gordon or Dirac Lagrangian) and to consider global phase transformations of
the wave function,
\beq \label{globalphase}
\psi_\tau(x) = e^{i \tau} \psi(x) \:, \qquad x_\tau = x \:.
\eeq
Then the symmetry condition~\eqref{symm} is satisfied because the
Lagrangian depends only on the modulus of~$\psi$. 
The corresponding Noether current is the probability current,
giving rise to {\em{current conservation}}.
We remark that, if the quantum mechanical wave function is coupled to an electromagnetic field,
then this current coincides, up to a multiplicative constant, with the electromagnetic current of the
particle. Therefore, the conservation law can also be interpreted as the {\em{conservation of electric
charge}}. The second application is to consider translations in space-time, i.e.
\[ \psi_\tau(x) = \psi(x) \:, \qquad x_\tau = x + \tau v \]
with a fixed vector~$v \in \scrM$. In this case, the symmetry condition~\eqref{symm}
is satisfied if we assume that~$\L=\L(\phi, \phi_{,j})$ does not depend explicitly on~$x$.
After a suitable symmetrization procedure (see~\cite[\S32 and~\S94]{landau2} or the
systematic treatment in~\cite{forger+roemer}),
the corresponding Noether current can be written as
\[ J^k = T^{kj} v_j \:, \]
where~$T_{jk}$ is the energy-momentum tensor.
Noether's theorem yields the {\em{conservation of energy and momentum}}.

Noether's theorem also applies in curved space-time. In this case, the Lagrangian
involves the Lorentzian metric. As a consequence, the symmetry condition~\eqref{symm}
implies that the metric must be invariant under the variation~$x_\tau$.
This is made precise by the notion of a {\em{Killing field}}~$K$, being a
vector field which satisfies the Killing equation
\[ \nabla_i K_j = - \nabla_j K_i \]
(see for example~\cite[Section~2.6]{hawking+ellis} or~\cite[Section~1.9]{straumann}).
If space-time admits a Killing field~$K$, the corresponding Noether current is
most conveniently constructed as follows. 
As a consequence of the Einstein equations, the energy-momentum tensor is divergence-free,
\[ \nabla_j T^{jk} = 0 \:. \]
This by itself does not give rise to conserved quantities because the Gau{\ss} divergence theorem
only applies to vector fields, but not to tensor fields.
However, a direct computation shows that contracting the energy-momentum tensor
with the Killing field,
\[ J^k := T^{kj} K_j \:, \]
gives rise to a divergence-free vector field (see~\cite[Section~3.2]{hawking+ellis}
or~\cite[Section~2.4]{straumann}). Now integration again gives the conservation law~\eqref{conserve}.

\subsection{Causal Variational Principles in the Compact Setting} \label{secintroc}
We now introduce the setting of causal variational principles in the compact case,
slightly generalizing the presentation in~\cite[Section~1.2]{support}.
Let~$\F$ be a smooth {\em{compact}} manifold and~$\L \in C^{0,1}(\F \times \F, \R^+_0)$
a non-negative Lipschitz-continuous function which is symmetric, i.e.
\beq \label{symmL}
\L(x,y) = \L(y,x) \qquad \text{for all~$x,y \in \F$}\:.
\eeq
The {\em{causal variational principle}} is to minimize the action~$\Sact$ defined by
\beq \label{Sdefcompact}
\Sact(\rho) = \iint_{\F \times \F} \L(x,y)\: d\rho(x)\: d\rho(y)
\eeq
under variations of~$\rho$ in the class of (positive) normalized regular Borel measures.
The existence of minimizers follows immediately from abstract compactness arguments
(see~\cite[Section~1.2]{continuum}).

In what follows, we let~$\rho$ be a given minimizing measure, referred to as the {\em{universal measure}}.
The resulting EL equations are derived in~\cite[Section~3.1]{support}. For the sake of self-consistency,
we now state them and repeat the proof.
\begin{Lemma} {\bf{(Euler-Lagrange equations)}} \label{lemmaEL}
Let~$\rho$ be a minimizing measure of the causal variational principle~\eqref{Sdefcompact}.
Then the function~$\ell \in C^{0,1}(\F)$ defined by
\beq
\ell(x) = \int_\F \L(x,y)\: d\rho(y) \label{ldef}
\eeq
is minimal on the support of~$\rho$,
\beq \label{EL1}
\ell|_{\supp \rho} \,\equiv\, \inf_\F \ell \:.
\eeq
\end{Lemma}
\Proof Carrying out one of the integrals, one sees that
\beq \label{Sl}
\Sact(\rho) = \iint_{\F \times \F} \L(x,y)\, d\rho(x)\: d\rho(y) = \int_\F \ell\: d\rho \:.
\eeq
Since~$\ell$ is continuous and~$\F$ is compact, there clearly is~$y \in \F$ with
\[ \ell(y) = \inf_\F \ell \:. \]
We consider for~$\tau \in [0,1]$ the family of normalized regular Borel measures
\[ \tilde{\rho}_\tau = (1-\tau)\, \rho + \tau \, \delta_y \:, \]
where~$\delta_y$ denotes the Dirac measure supported at~$y$. Applying this formula in~\eqref{Sdefcompact}
and differentiating, we obtain for the first variation
\[ \delta \Sact := \lim_{t \searrow 0} \frac{\Sact \big(\tilde{\rho}_\tau \big) - \Sact\big(\tilde{\rho}_0 \big)}{\tau}
= -2 \Sact(\rho) + 2 \ell(y)\:. \]
Since~$\rho$ is a minimizer, $\delta \Sact$ is non-negative. Hence
\[ \inf_\F \ell = \ell(y) \:\geq\: \Sact(\rho) \overset{\eqref{Sl}}{=} \int_\F \ell\: d\rho \:. \]
It follows that~$\ell$ is constant on the support of~$\rho$, giving the result.
\QED

The physical picture is that the universal measure gives rise to a space-time and also induces all
the objects therein. In the compact setting considered here, one only obtains space-time
endowed with a causal structure in the following way.
Space-time is defined as the support of the universal measure,
\[ \text{\em{space-time}} \qquad M:= \supp \rho \:. \]
For a space-time point~$x \in M$,
we define the open {\em{light cone}} ${\mathcal{I}}(x)$ and the
closed light cone~${\mathcal{J}}(x)$ by
\[ {\mathcal{I}}(x) = \{ y \in M \:|\: \L(x,y) > 0 \} \qquad \text{and} \qquad
 {\mathcal{J}}(x) = \overline{{\mathcal{I}}(x)}\:. \]
This makes it possible to define a {\em{causal structure}} on space-time
by saying that two space-time points~$x, y \in M$ are
{\em{timelike}} separated if $\L(x,y)>0$ and {\em{spacelike}} separated if~$\L(x,y)=0$.
We remark that, in the setting of causal fermion systems, these notions
indeed agree with the usual notion of causality in Minkowski space or on a globally
hyperbolic manifold (we refer the interested reader to~\cite{dice2014} or~\cite{cfs}).

\subsection{The Concept of Surface Layer Integrals} \label{secsli}
It is not at all obvious how the classical Noether theorem should be generalized
to causal variational principles. First, the mathematical structure of the
EL equations~\eqref{EL1} is completely different from that
of the classical EL equations~\eqref{ELclass}.
Moreover, for writing down surface integrals as in~\eqref{conserve}
one needs structures like the Lorentzian metric as well as the normal to a hypersurface
and the induced volume measure thereon. All these structures are
not available in the setting of causal variational principles.
Therefore, it is a-priori not clear how conservation laws should be stated.

\begin{figure}
\psscalebox{1.0 1.0} 
{
\begin{pspicture}(0,-1.511712)(10.629875,1.511712)
\definecolor{colour0}{rgb}{0.8,0.8,0.8}
\definecolor{colour1}{rgb}{0.6,0.6,0.6}
\pspolygon[linecolor=black, linewidth=0.002, fillstyle=solid,fillcolor=colour0](6.4146066,0.82162136)(6.739051,0.7238436)(6.98794,0.68384355)(7.312384,0.66162133)(7.54794,0.67939913)(7.912384,0.7593991)(8.299051,0.8705102)(8.676828,0.94162136)(9.010162,0.9549547)(9.312385,0.9371769)(9.690162,0.8571769)(10.036829,0.7371769)(10.365718,0.608288)(10.614607,0.42162135)(10.614607,-0.37837866)(6.4146066,-0.37837866)
\pspolygon[linecolor=black, linewidth=0.002, fillstyle=solid,fillcolor=colour1](6.4146066,1.2216214)(6.579051,1.1616213)(6.770162,1.1127324)(6.921273,1.0905102)(7.103495,1.0816213)(7.339051,1.0549546)(7.530162,1.0638436)(7.721273,1.0993991)(7.8857174,1.1393992)(8.10794,1.2060658)(8.299051,1.2549547)(8.512384,1.3038436)(8.694607,1.3260658)(8.890162,1.3305103)(9.081273,1.3393991)(9.379051,1.3216213)(9.659051,1.2593992)(9.9746065,1.1705103)(10.26794,1.0460658)(10.459051,0.94384354)(10.614607,0.82162136)(10.610162,0.028288014)(10.414606,0.1660658)(10.22794,0.26828802)(10.010162,0.37051025)(9.663495,0.47273245)(9.356829,0.53051025)(9.054606,0.548288)(8.814607,0.54384357)(8.58794,0.5171769)(8.387939,0.48162135)(8.22794,0.44162133)(7.90794,0.34828803)(7.6946063,0.29939914)(7.485718,0.26828802)(7.272384,0.26828802)(7.02794,0.28162134)(6.82794,0.3171769)(6.676829,0.35273245)(6.543495,0.38828802)(6.4146066,0.42162135)
\pspolygon[linecolor=black, linewidth=0.002, fillstyle=solid,fillcolor=colour0](0.014606438,0.82162136)(0.3390509,0.7238436)(0.5879398,0.68384355)(0.9123842,0.66162133)(1.1479398,0.67939913)(1.5123842,0.7593991)(1.8990508,0.8705102)(2.2768288,0.94162136)(2.610162,0.9549547)(2.9123843,0.9371769)(3.290162,0.8571769)(3.6368287,0.7371769)(3.9657176,0.608288)(4.2146063,0.42162135)(4.2146063,-0.37837866)(0.014606438,-0.37837866)
\psbezier[linecolor=black, linewidth=0.04](6.4057174,0.8260658)(7.6346064,0.45939913)(7.8634953,0.8349547)(8.636828,0.92828804)(9.410162,1.0216213)(10.165717,0.7927325)(10.614607,0.42162135)
\psbezier[linecolor=black, linewidth=0.04](0.005717549,0.8260658)(1.2346064,0.45939913)(1.4634954,0.8349547)(2.2368286,0.92828804)(3.0101619,1.0216213)(3.7657175,0.7927325)(4.2146063,0.42162135)
\rput[bl](2.0101619,0.050510235){$\Omega$}
\rput[bl](8.759051,0.0016213481){\normalsize{$\Omega$}}
\psline[linecolor=black, linewidth=0.04, arrowsize=0.09300000000000001cm 1.0,arrowlength=1.7,arrowinset=0.3]{->}(1.9434953,0.85495466)(1.8057176,1.6193991)
\rput[bl](2.0946064,1.1705103){$\nu$}
\psbezier[linecolor=black, linewidth=0.02](6.4146066,0.42384356)(7.6434956,0.057176903)(7.872384,0.43273246)(8.645718,0.52606577)(9.419051,0.61939913)(10.174606,0.39051023)(10.623495,0.019399125)
\psbezier[linecolor=black, linewidth=0.02](6.410162,1.2193991)(7.639051,0.8527325)(7.86794,1.228288)(8.6412735,1.3216213)(9.414606,1.4149547)(10.170162,1.1860658)(10.619051,0.8149547)
\rput[bl](8.499051,0.9993991){\normalsize{$y$}}
\rput[bl](7.8657174,0.49273247){\normalsize{$x$}}
\psdots[linecolor=black, dotsize=0.06](8.170162,0.65273243)
\psdots[linecolor=black, dotsize=0.06](8.796828,1.1327325)
\psline[linecolor=black, linewidth=0.02](6.1146064,1.2216214)(6.103495,0.82162136)
\rput[bl](5.736829,0.8993991){\normalsize{$\delta$}}
\rput[bl](3.6146064,0.888288){$\scrN$}
\rput[bl](1.1146064,-1.4117119){$\displaystyle \int_\scrN \cdots\, d\mu_\scrN$}
\rput[bl](5.7146063,-1.511712){$\displaystyle \int_\Omega d\rho(x) \int_{M \setminus \Omega} d\rho(y)\: \cdots\:\L(x,y)$}
\psline[linecolor=black, linewidth=0.02](6.0146065,1.2216214)(6.2146063,1.2216214)
\psline[linecolor=black, linewidth=0.02](6.0146065,0.82162136)(6.2146063,0.82162136)
\end{pspicture}
}
\caption{A surface integral and a corresponding surface layer integral.}
\label{fignoether1}
\end{figure}
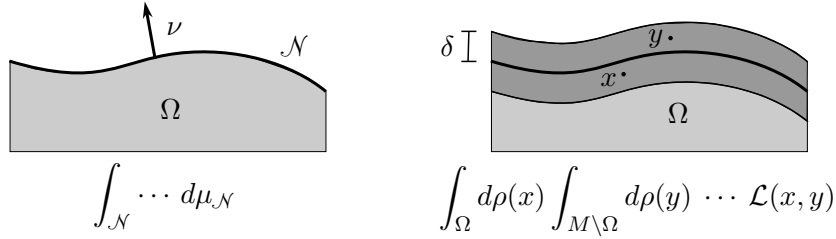

The first task is to introduce an analog of the surface integral in~\eqref{conserve}.
The only objects to our disposal are the Lagrangian~$\L(x,y)$ and the universal measure~$\rho$.
We make the assumption that the Lagrangian is of {\em{short range}} in the following sense.
We let~$d \in C^0(M \times M, \R^+_0)$ be a distance function on~$M$
(since~$M$ is compact, any two such distance functions are equivalent). The assumption
of short range means that~$\L$ vanishes on distances larger than~$\delta$, i.e.
\beq \label{shortrange}
d(x,y) > \delta \quad \Longrightarrow \quad \L(x,y) = 0
\eeq
Then a double integral of the form
\beq \label{intdouble}
\int_\Omega \bigg( \int_{M \setminus \Omega} \cdots\: \L(x,y)\: d\rho(y) \bigg)\, d\rho(x)
\eeq
only involves pairs~$(x,y)$ of distance at most~$\delta$,
where~$x$ is in~$\Omega$ and~$y$ is in the complement~$M \setminus \Omega$.
Thus the integral only involves points in a layer around the boundary of~$\Omega$
of width~$\delta$, i.e.
\[ x, y \in B_\delta \big(\partial \Omega \big) \:. \]
Therefore, a double integral of the form~\eqref{intdouble} can be regarded as an approximation
of a surface integral on the length scale~$\delta$, as shown in Figure~\ref{fignoether1}.
We refer to integrals of the form~\eqref{intdouble} as {\em{surface layer integrals}}.
In the setting of causal variational principles, they take the role of surface integrals
in Lorentzian geometry.
Our strategy is to find expressions for the integrand ``\ldots'' in~\eqref{intdouble} such that the
surface layer integral vanishes. Choosing~$\Omega$ as a space-time region such that~$\delta \Omega$
has two connected components~$\scrN_1$ and~$\scrN_2$, one then obtains
a conservation law similar to~\eqref{conserve}, with the surface integrals replaced by
corresponding surface layer integrals.

We remark for clarity that the correspondence between surface integrals and surface layer integrals
could be made mathematically precise by taking the limit~$\delta \searrow 0$.
However, this would make it necessary to consider a family of Lagrangians~$\L_\delta$
together with corresponding minimizers~$\rho_\delta$.
This seems an interesting technical problem for the future.
For our purposes, it suffices to identify the surface layer integrals~\eqref{intdouble}
as the objects which replace the usual surface integrals.

We finally remark that, in the physical setting of causal fermion systems, the
condition of short range~\eqref{shortrange} will be replaced by the weaker requirement
that the main contribution to the double integral~\eqref{intdouble} comes from pairs
of points~$(x,y)$ whose distance is at most~$\delta$. This will be explained in 
detail in Section~\ref{seccurcor}, where will also identify the length scale~$\delta$
with the Compton scale (see the paragraph after after~\eqref{USymm20}). 

\section{Noether-Like Theorems in the Compact Setting} \label{seccompact}
We now derive Noether-like theorems in the compact setting.
We consider two different symmetries: symmetries of the Lagrangian (Theorem~\ref{thmsymmlag})
and symmetries of the universal measure (Theorem~\ref{thmsymmum}).
In Section~\ref{secsymmgis}, these symmetries will be combined in
the notion of generalized integrated symmetries (Theorem~\ref{thmsymmgis}).

\subsection{Symmetries of the Lagrangian} \label{secsymmlag}
The assumption~\eqref{symm} can be understood as a symmetry condition
for the Lagrangian. We now want to impose a similar symmetry condition for the
Lagrangian~$\L(x,y)$ of a causal variational principle. The most obvious method would be to
consider a one-parameter group of diffeomorphisms~$\Phi_\tau$,
\beq \label{try}
\Phi : \R \times \F \rightarrow \F \qquad \text{with} \qquad
\Phi_\tau \Phi_{\tau'} = \Phi_{\tau+\tau'}
\eeq
and to impose that~$\L$ be invariant under these diffeomorphisms in the sense that
\beq \label{symmF1}
\L(x,y) = \L \big( \Phi_\tau(x), \Phi_\tau(y) \big) \qquad \text{for all~$\tau \in \R$
and~$x, y \in \F$\:.}
\eeq
However, this condition is unnecessarily strong for two reasons. First, it suffices to consider families which
are defined locally for~$\tau \in (-\tau_{\max}, \tau_{\max})$. Second, the mapping~$\Phi$ does not
need to be defined on all of~$\F$. Instead, it is more appropriate to impose the symmetry condition
only on space-time~$M \subset \F$.
This leads us to  consider instead of~\eqref{try} a mapping
\beq \label{Phidef}
\Phi : (-\tau_{\max}, \tau_{\max}) \times M \rightarrow \F
\qquad \text{with} \qquad \Phi(0,.) = \1 \:.
\eeq
We also write~$\Phi_\tau(x) \equiv \Phi(\tau,x)$ and refer to~$\Phi_\tau$
as a {\bf{variation}} of~$M$ in~$\F$.
Next, we need to specify what we mean by ``smoothness'' of this variation.
This is a subtle point because in view of the results in~\cite{support},
the universal measure does not need to be smooth (in the sense that it cannot in general be written as
a smooth function times the Lebesgue measure), and therefore the function~$\ell$
will in general only be Lipschitz continuous.
Our Noether-like theorems require only that the function~$\ell$ be differentiable in the
direction of the variations:
\begin{Def} \label{defsymm} A variation~$\Phi_\tau$ of the form~\eqref{Phidef} is
{\bf{continuously differentiable}} if the composition
\[ \ell \circ \Phi \::\: (-\tau_{\max}, \tau_{\max}) \times M \rightarrow \R \]
is continuous and if its partial derivative~$\partial_\tau (\ell \circ \Phi)$ exists
and is continuous.
\end{Def}
The next question is how to adapt the symmetry condition~\eqref{symmF1} to the mapping~$\Phi$
defined only on~$(-\tau_{\max}, \tau_{\max}) \times M$.
This is not obvious because setting~$\tilde{x} = \Phi_\tau(x)$ and using the group property,
the condition~\eqref{symmF1} can be written equivalently as
\beq \label{symmF2}
 \L \big( \Phi_{-\tau}(\tilde{x}), y \big) = \L \big( \tilde{x},\Phi_\tau(y) \big) \qquad \text{for all~$\tau \in \R$
 and~$\tilde{x}, y \in \F$\:.}
\eeq
But if we restrict attention to pairs~$x,y \in M$, the equations in~\eqref{symmF1} and~\eqref{symmF2}
are different. It turns out that the correct procedure is to work with the expression in~\eqref{symmF2}.

\begin{Def} \label{defsymmlagr} A variation~$\Phi_\tau$ of the form~\eqref{Phidef} is a {\bf{symmetry
of the Lagrangian}} if
\beq \label{symmlagr}
\L \big( x, \Phi_\tau(y) \big) = \L \big( \Phi_{-\tau}(x), y \big)
\qquad \text{for all~$\tau \in (-\tau_{\max}, \tau_{\max})$
and~$x, y \in M$\:.}
\eeq
\end{Def}

We now state our first Noether-like theorem.
\begin{Thm} \label{thmsymmlag} Let~$\rho$ be a minimizing measure and $\Phi_\tau$ a 
continuously differentiable symmetry of the Lagrangian. Then for any compact subset~$\Omega \subset M$,
we have
\beq
\label{conservationLagrEq}
\frac{d}{d\tau} \int_\Omega d\rho(x) \int_{M \setminus \Omega} d\rho(y)\:
\Big( \L \big( \Phi_\tau(x),y \big) - \L \big( \Phi_{-\tau}(x), y \big) \Big) \Big|_{\tau=0} = 0 \:.
\eeq
\end{Thm}
Before coming to the proof, we explain the connection to surface layer integrals.
To this end, let us assume that~$\Phi_\tau$ and the Lagrangian are differentiable in the sense that the derivatives
\beq \label{differentiable}
\frac{d}{d\tau} \Phi_\tau(x)\big|_{\tau=0} =: u(x) \qquad \text{and} \qquad
\frac{d}{d\tau} \L\big( \Phi_\tau(x),y \big) \big|_{\tau=0}
\eeq
exist for all~$x, y \in M$ and are continuous on~$M$ respectively~$M \times M$.
Then one may exchange differentiation and integration in~\eqref{conservationLagrEq}
and apply the chain rule to obtain
\[ \int_\Omega d\rho(x) \int_{M \setminus \Omega} d\rho(y)\: D_{u(x)} \L(x,y) = 0 \:, \]
where~$D_{u(x)}$ is the derivative in the direction of the vector field~$u(x)$.
This expression is a surface layer integral as in~\eqref{intdouble}.
In general, the derivatives in~\eqref{differentiable} need {\em{not}} exist, because
we merely imposed the weaker differentiability assumption of Definition~\ref{defsymm}.
In this case, the statement of the theorem implies that the derivative 
of the integral in~\eqref{conservationLagrEq} exists and vanishes.

\Proof[Proof of Theorem~\ref{thmsymmlag}.] We multiply~\eqref{symmlagr} by a bounded measurable function~$f$ on~$M$
and integrate. This gives
\begin{align*}
0 &= \iint_{M \times M} f(x)\, f(y)\: \Big(
\L \big( x, \Phi_\tau(y) \big) - \L \big( \Phi_{-\tau}(x), y \big) \Big)\, d\rho(x)\, d\rho(y) \\
&=\iint_{M \times M} f(x)\, f(y)\: \Big(
\L \big( \Phi_\tau(x),y \big) - \L \big( \Phi_{-\tau}(x), y \big) \big) \Big) \, d\rho(x)\, d\rho(y)\:,
\end{align*}
where in the last step we used the symmetry of the Lagrangian~\eqref{symmL}
and the symmetry of the integrand in~$x$ and~$y$.
We replace~$f(y)$ by~$1 - (1-f(y))$, multiply out and use the definition of~$\ell$, \eqref{ldef}.
We thus obtain
\begin{align*}
0 &=\int_M f(x)\: 
\Big(  \ell \big( \Phi_\tau(x)\big) - \ell\big( \Phi_{-\tau}(x)) \Big) \,d\rho(x) \\
&\quad -\iint_{M \times M} f(x)\, \big(1-f(y) \big)\:
\Big( \L \big( \Phi_\tau(x),y \big) - \L \big( \Phi_{-\tau}(x), y \big) \Big) \, d\rho(x)\, d\rho(y)\:.
\end{align*}
Choosing~$f$ as the characteristic function of~$\Omega$, we obtain the identity
\beq \begin{split}
\int_\Omega &d\rho(x) \: \int_{M \setminus \Omega} d\rho(y)\:
\Big( \L \big( \Phi_\tau(x), y \big) - \L \big(\Phi_{-\tau}(x),  y \big) \Big) \\
&= \int_\Omega \Big( \ell \big( \Phi_\tau(x) \big)  - \ell \big( \Phi_{-\tau}(x) \big) \Big)\: d\rho(x) \:.
\end{split} \label{form}
\eeq
Using that~$\ell(\Phi_\tau(x))$ is continuously differentiable (see Definition~\ref{defsymm})
and that~$\Omega$ is compact, we conclude that
the right side of this equation is differentiable at~$\tau=0$. Moreover,
we are allowed to exchange the $\tau$-differentiation with integration.
The EL equations~\eqref{EL1} imply that
\beq \label{lfirst}
\frac{d}{d\tau} \ell \big( \Phi_{\tau}(x) \big) \Big|_{\tau=0} = 0 =
\frac{d}{d\tau} \ell \big( \Phi_{-\tau}(x) \big) \Big|_{\tau=0} \:.
\eeq
Hence the right side of~\eqref{form} is differentiable at~$\tau=0$,
and the derivative vanishes. This gives the result.
\QED

\subsection{Symmetries of the Universal Measure} \label{secsymmum}
We now prove a conservation law for a different type of symmetry.

\begin{Def} \label{defsymmrho} A variation~$\Phi_\tau$
of the form~\eqref{Phidef} is a {\bf{symmetry of the universal measure}} if
\beq \label{symmrho}
(\Phi_\tau)_* \rho = \rho
\qquad \text{for all~$\tau \in (-\tau_{\max}, \tau_{\max})$\:.}
\eeq
\end{Def} \noindent
Here~$(\Phi_\tau)_* \rho$ is the push-forward measure
(defined by~$((\Phi_\tau)_* \rho)(\Omega) := \rho(\Phi_\tau^{-1}(\Omega))$).

\begin{Thm} \label{thmsymmum} Let~$\rho$ be a minimizing measure and $\Phi_\tau$  be a
continuously differentiable symmetry of the universal measure. Then for any compact subset~$\Omega \subset M$,
\[ \frac{d}{d\tau} \int_\Omega d\rho(x) \int_{M \setminus \Omega} d\rho(y)\: \Big( 
\L \big( \Phi_\tau(x), y \big)  - \L \big( x, \Phi_\tau(y) \big) \Big) \Big|_{\tau=0} = 0 \:. \]
\end{Thm}
\Proof We again let~$f$ be a bounded measurable function on~$M$. Then, by symmetry
in~$x$ and~$y$,
\[ \iint_{M \times M} f(x)\,f(y)\: \Big( \L \big( \Phi_\tau(x), y \big) -\L \big( x, \Phi_\tau(y) \big) \Big)
\: d\rho(x)\, d\rho(y) = 0 \:. \]
We replace~$f(y)$ by~$1 - (1-f(y))$ and multiply out. The double integrals which do not involve~$f(y)$
can be simplified as follows,
\begin{align*}
&\iint_{M \times M} f(x)\, \L \big( \Phi_\tau(x), y \big) \: d\rho(x)\, d\rho(y) = \int_M f(x)\:\ell \big( \Phi_\tau(x) \big) 
\: d\rho(x) \\
&\iint_{M \times M} f(x)\, \L \big( x, \Phi_\tau(y) \big) \: d\rho(x)\, d\rho(y) =
\iint_{\F \times \F} f(x)\, \L \big( x, \Phi_\tau(y) \big) \: d\rho(x)\, d\rho(y) \\
&=\iint_{\F \times \F} f(x)\, \L(x,y) \: d\rho(x)\, d \big((\Phi_\tau)_*\rho \big)(y) 
\overset{(\star)}{=} \iint_{\F \times \F} f(x)\, \L(x,y) \: d\rho(x)\, d \rho (y) \\
&= \iint_{M \times M} f(x)\, \L(x,y) \: d\rho(x)\, d \rho (y)
= \int_M f(x)\, \ell(x)\, d\rho(x) \:,
\end{align*}
where in~($\star$) we used the symmetry assumption~\eqref{symmrho}.
We thus obtain
\begin{align*}
0 &= -\iint_{M \times M} f(x)\,\big(1-f(y) \big)\: \Big( \L \big( \Phi_\tau(x), y \big) 
- \L \big( x, \Phi_\tau(y) \big) \Big)\: d\rho(x)\, d\rho(y) \\
&\quad + \int_M f(x)\, \Big( \ell \big( \Phi_\tau(x) \big) - \ell(x) \Big) \: d\rho(x)\:.
\end{align*}
Choosing~$f$ as the characteristic function of~$\Omega$ gives
\[ \int_\Omega d\rho(x) \int_{M \setminus \Omega} d\rho(y)\: \Big( \L \big( \Phi_\tau(x), y \big) -
\L \big( x, \Phi_\tau(y) \big) \Big)
= \int_\Omega \Big(  \ell \big( \Phi_\tau(x) \big) - \ell(x) \Big) \: d\rho(x) \:. \]
Now the $\tau$-derivative can be computed just as in the proof of Theorem~\ref{thmsymmlag}.
\QED

\subsection{Generalized Integrated Symmetries} \label{secsymmgis}
We now combine the symmetries of the previous sections
in the notion of ``generalized integrated symmetries.''
Our method is based on the following simple but useful identity.
\begin{Prp} \label{prpuseful} Let~$\Phi_\tau$ be a variation of the form~\eqref{Phidef}. Then
\begin{align}
\int_M d\rho&(x) \int_\Omega d\rho(y) \Big( \L\big( \Phi_\tau(x), y\big) - \L(x,y) \Big) \label{id1} \\
=\:& \int_\Omega \Big( \ell \big( \Phi_\tau(x)\big) - \ell(x) \Big)\, d\rho(x) \label{id2} \\
&- \int_\Omega d\rho(x) \int_{M \setminus \Omega} d\rho(y) \:\Big( \L\big( \Phi_\tau(x), y\big) -
\L\big( x, \Phi_\tau(y) \big) \Big) \:. \label{id3}
\end{align}
\end{Prp}
\Proof We rewrite the integration domains as follows,
\begin{align}
\int_M & d\rho(x) \int_\Omega d\rho(y) \Big( \L\big( \Phi_\tau(x), y\big) - \L(x,y) \Big) \notag \\
&= \int_\Omega d\rho(x) \int_\Omega d\rho(y) \Big( \L\big( \Phi_\tau(x), y\big) - \L(x,y) \Big) \notag \\
&\quad+ \int_{M \setminus \Omega} d\rho(x) \int_\Omega d\rho(y) \Big( \L\big( \Phi_\tau(x), y\big) - \L(x,y) \Big) \notag \\
&= \int_\Omega d\rho(x) \int_M d\rho(y) \Big( \L\big( \Phi_\tau(x), y\big) - \L(x,y) \Big) \label{int1} \\
&\quad- \int_\Omega d\rho(x) \int_{M \setminus \Omega} d\rho(y) \Big( \L\big( \Phi_\tau(x), y\big) - \L(x,y) \Big) \\
&\quad+ \int_{M \setminus \Omega} d\rho(x) \int_\Omega d\rho(y) \Big( \L\big( \Phi_\tau(x), y\big) - \L(x,y) \Big) \:.
\label{int3}
\end{align}
In~\eqref{int1} we can carry out the $y$-integration using~\eqref{ldef}.
In~\eqref{int3} we exchange the integrals and use that the Lagrangian is symmetric in its
two arguments~\eqref{symmL}. This gives the result.
\QED

Note that the term~\eqref{id3} is a boundary layer integral. The term~\eqref{id2}, on the
other hand, only involves~$\ell$, and therefore its first variation vanishes in view
of the EL equations~\eqref{EL1}. We thus obtain a conservation law, provided that
the term~\eqref{id1} vanishes. This motivates the following definition.

\begin{Def} \label{defgis}
A variation~$\Phi_\tau$ of the form~\eqref{Phidef} is a {\bf{generalized integrated symmetry}} in
the space-time region~$\Omega \subset M$ if
\beq \label{symmgis}
\int_M d\rho(x) \int_\Omega d\rho(y) \Big( \L\big( \Phi_\tau(x), y\big) - \L(x,y) \Big)  = 0 \:.
\eeq
\end{Def}
This notion of symmetry indeed generalizes our previous notions of symmetry
(see Definitions~\ref{defsymmlagr} and~\ref{defsymmrho}) in the sense that
symmetries of the Lagrangian and of the universal measure imply that~\eqref{symmgis}
holds for first variations. Namely, if~$\Phi_\tau$ is a symmetry of the universal measure,
we can use~\eqref{symmrho} to obtain
\beq \begin{split}
&\int_M d\rho(x) \int_\Omega d\rho(y) \Big( \L\big( \Phi_\tau(x), y\big) - \L(x,y) \Big) \\
&= \int_\F d \big( (\Phi_{\tau})_* \rho \big)(x) \int_\Omega d\rho(y) \: \L(x,y)
- \int_\F d\rho(x) \int_\Omega d\rho(y) \:\L(x,y) = 0 \:. \label{calc1}
\end{split}
\eeq
Likewise, if~$\Phi_\tau$ is a symmetry of the Lagrangian, we can apply~\eqref{symmlagr}. This gives
the identity
\beq \begin{split}
&\int_M d\rho(x) \int_\Omega d\rho(y) \Big( \L\big( \Phi_\tau(x), y\big) - \L(x,y) \Big) \\
&= \int_M d\rho(x) \int_\Omega d\rho(y) \Big( \L\big( x, \Phi_{-\tau}(y) \big) - \L(x,y) \Big) \\
&= \int_\Omega \Big( \ell \big( \Phi_{-\tau}(y) \big) - \ell(y) \Big)\: d\rho(y) \:, \label{calc2}
\end{split}
\eeq
whose first variation vanishes in view of~\eqref{lfirst}.

Combining Definition~\ref{defgis} with Proposition~\ref{prpuseful} immediately gives the following result.
\begin{Thm} \label{thmsymmgis}
Let~$\rho$ be a minimizing measure and $\Phi_\tau$ a 
continuously differentiable generalized integrated symmetry (see Definition~\ref{defgis}).
Then for any compact subset~$\Omega \subset M$,
\beq
\frac{d}{d\tau} \int_\Omega d\rho(x) \int_{M \setminus \Omega} d\rho(y) \:\Big( \L\big( \Phi_\tau(x), y\big) -
\L\big( x, \Phi_\tau(y) \big) \Big) \Big|_{\tau=0} = 0 \:.
\eeq
\end{Thm} \noindent
In view of~\eqref{calc1} and~\eqref{calc2},
the previous conservation laws of Theorems~\ref{thmsymmlag} and~\ref{thmsymmum}
are immediate corollaries of this theorem.

\section{The Setting of Causal Fermion Systems} \label{seccfs}
We now turn attention to the setting of causal fermion systems.
After a short review of the mathematical framework and the Euler-Lagrange equations
(Section~\ref{seccfsbasic}), we prove Noether-like theorems (Section~\ref{seccfsnoether}).
The reader interested in a more detailed introduction to causal fermion systems
is referred to the introductory chapter in~\cite{cfs}.
A non-technical introduction is given in~\cite{dice2014}.

\subsection{Basic Definitions and the Euler-Lagrange Equations} \label{seccfsbasic}
\begin{Def} \label{defparticle} {\bf{(causal fermion system)}} {\em{
Given a separable complex Hilbert space $\H$ with scalar product~$\la .|. \ra_\H$
and a parameter~$n \in \N$ (the {\em{``spin dimension''}}), we let~$\F \subset \Lin(\H)$ be the set of all
self-adjoint operators on~$\H$ of finite rank, which (counting multiplicities) have
at most~$n$ positive and at most~$n$ negative eigenvalues. On~$\F$ we are given
a positive measure~$\rho$ (defined on a $\sigma$-algebra of subsets of~$\F$), the so-called
{\em{universal measure}}. We refer to~$(\H, \F, \rho)$ as a {\em{causal fermion system}}.
}}
\end{Def}

We next introduce the causal action principle. For any~$x, y \in \F$, the product~$x y$ is an operator
of rank at most~$2n$. We denote its non-trivial eigenvalues (counting algebraic multiplicities)
by~$\lambda^{xy}_1, \ldots, \lambda^{xy}_{2n} \in \C$.
We introduce the {\em{spectral weight}}~$| \,.\, |$ of an operator as the sum of the absolute values
of its eigenvalues. In particular, the spectral weight of the operator
products~$xy$ and~$(xy)^2$ is defined by
\[ |xy| = \sum_{i=1}^{2n} \big| \lambda^{xy}_i \big|
\qquad \text{and} \qquad \big| (xy)^2 \big| = \sum_{i=1}^{2n} \big| \lambda^{xy}_i \big|^2 \:. \]
We introduce the Lagrangian and the action by
\begin{align}
\text{\em{Lagrangian:}} && \L(x,y) &= \big| (xy)^2 \big| - \frac{1}{2n}\: |xy|^2 \label{Lagrange} \\
\text{\em{action:}} && \Sact(\rho) &= \iint_{\F \times \F} \L(x,y)\: d\rho(x)\, d\rho(y) \:. \label{Sdef}
\end{align}
The {\em{causal action principle}} is to minimize~$\Sact$ by varying the universal measure
under the following constraints:
\begin{align}
\text{\em{volume constraint:}} && \rho(\F) = \text{const} > 0 \quad\;\; & \label{volconstraint} \\
\text{\em{trace constraint:}} && \int_\F \tr(x)\: d\rho(x) = \text{const} \neq 0 & \label{trconstraint} \\
\text{\em{boundedness constraint:}} && \T(\rho) := \iint_{\F \times \F} |xy|^2\: d\rho(x)\, d\rho(y) &\leq C \:, \label{Tdef}
\end{align}
where~$C$ is a given parameter (and~$\tr$ denotes the trace of linear operators on~$\H$).

\subsubsection{The finite-dimensional setting}
If~$\H$ is {\em{finite-dimensional}} and~$\rho$ has {\em{finite total volume}}, the existence of
minimizers is proven in~\cite{continuum}, and the corresponding EL equations
are derived in~\cite{lagrange}. We now recall a few of these results.
Under the above assumptions, on~$\F$ one considers the topology induced by the
operator norm
\beq \label{supnorm}
\|A\| := \sup \big\{ \|A u \|_\H \text{ with } \| u \|_\H = 1 \big\} \:.
\eeq
In this topology, the Lagrangian as well as the integrands in~\eqref{trconstraint}
and~\eqref{Tdef} are continuous. We vary~$\rho$ within the class of bounded Borel measures of~$\F$.
The existence of minimizers of the action~\eqref{Sdef} under the constraints~\eqref{volconstraint}--\eqref{Tdef} is proven in~\cite[Theorem~2.1]{continuum}.
For our purposes, the resulting EL equations are most conveniently
stated as follows (for a heuristic derivation see the introduction in~\cite{lagrange}).

\begin{Thm} \label{thm3} Assume that~$\rho$ is a minimizer of the causal
action principle for~$C$ so large that
\beq
C > \inf \big\{ \T(\mu) \:|\: \text{$\mu$ satisfies~\eqref{volconstraint}  and~\eqref{trconstraint}} \big\} \:. \label{Cbound}
\eeq
Moreover, assume that one of the following two technical assumptions hold:
\begin{enumerate}
\item[\rm{(i)}] The boundedness constraint is satisfied with a strict inequality,
\beq
\T(\rho) < C \:. \label{Cbound2}
\eeq
\item[\rm{(ii)}] The minimizer is regular in the sense of~\cite[Definition~3.12]{lagrange}.
\end{enumerate}
Then for a suitable choice of Lagrange multipliers~$\lambda, \kappa \in \R$,
the measure~$\rho$ is supported on the intersection of the level sets
\beq \label{hyper}
\Phi_1(x) = -4 \Sact(\rho)  \qquad \text{and} \qquad
\Phi_2(x) = 2 \Sact(\rho) \:,
\eeq
where
\beq
\Phi_1(x) := -\lambda \tr(x) \:,\qquad
\Phi_2(x) := 2 \int_\F \L_\kappa(x,y) \,d\rho(y) \label{Phi2def}
\eeq
and
\[ \L_\kappa(x,y) := \L(x,y) + \kappa \, |xy|^2 \:. \]
Moreover, the function
\[ \Phi(x) := \Phi_1 + \Phi_2 \]
is minimal on the support of~$\rho$, i.e.
\beq \label{Phiminimal}
\Phi|_{\supp \rho} = \inf_\F \Phi \:.
\eeq
\end{Thm}
\Proof We first apply~\cite[Theorem~1.3]{lagrange} to the causal variational principle
with trace constraint in the case~$\T(\rho)<C$. This yields that~$\rho$ is supported on the
intersection of the level sets~\eqref{hyper}. Moreover, this theorem implies that~$\Phi|_{\supp \rho}
= -2 \Sact(\rho)$.
The minimality~\eqref{Phiminimal} is proven in~\cite[Theorem~3.13]{lagrange}, noting
that the regularity condition of~\cite[Definition~3.12]{lagrange} is automatically satisfied
if the trace constraint is considered and if~\eqref{Cbound2} holds.
\QED
We remark for clarity that the inequality~\eqref{Cbound} can always be arranged
by choosing~$C$ sufficiently large.
The assumptions~(i) or~(ii) are needed in order for the Lagrange multiplier method
to be applicable. The basic difficulty comes about because the set of positive Borel
measures is not a vector space, but only a convex set.
Moreover, one must make sure that the constraints describe locally
a Banach submanifold. We refer the reader interested in the technical details to the
paper~\cite{lagrange}. In what follows, we take the assumptions~(i) or~(ii) for granted.

For the derivation of our conservation laws, we only need a weaker version of the
EL equations~\eqref{Phiminimal}. Namely, it suffices to assume that the function~$\Phi$ is constant on the
support of~$\rho$,
\beq \label{Phiminimala}
\Phi|_{\supp \rho} = \text{const} \:,
\eeq
and that the support of~$\rho$ is a {\em{local}} minimum in the sense that every~$x \in \supp \rho$ has a
neighborhood~$U(x) \subset \F$ such that
\beq \label{Phiminimalb}
\Phi(x) = \inf_{U(x)} \Phi \:.
\eeq
We subsume~\eqref{Phiminimala} and~\eqref{Phiminimalb} by saying that~$\rho$ is a
{\em{local minimizer}} of the causal action principle.
Working with local minimizers is also preferable because
the regularized Dirac sea configurations to be considered in the examples
of Sections~\ref{seccurcor} and~\ref{CorrDiracEM} are known to satisfy~\eqref{Phiminimala}
and~\eqref{Phiminimalb} in the continuum limit, but
but they are not global minimizers of the causal action principle
(for a detailed discussion of this point in the connection to microscopic mixing
and second-quantized bosonic fields we refer to~\cite[\S1.5.3]{cfs}).

\subsubsection{The infinite-dimensional setting}
We next consider the case that~$\H$ is infinite-dimensional or the total volume~$\rho(\F)$ is infinite.
First, a scaling argument shows that in the case~$\rho(\F)=\infty$ and~$\dim \H<\infty$, the
action is infinite for all measures satisfying the constraints, so that the variational principle is not sensible.
Similarly, if~$\rho(\F)<\infty$ and~$\dim \H=\infty$, the infimum of the action is zero,
but this infimum is not attained (for details see~\cite[Exercise~1.2]{cfs}). Therefore, the only interesting case is the
{\em{infinite-dimensional}} setting
when~$\rho(\F)=\infty$ and~$\dim \H=\infty$. In this setting, the causal action principle
makes mathematical sense if the volume constraint~\eqref{volconstraint}
is implemented by demanding that the variations~$(\rho(\tau))_{\tau \in (-\tau_{\max}, \tau_{\max})}$
should for all~$\tau, \tau' \in (-\tau_{\max}, \tau_{\max})$ satisfy the conditions
\[ \big| \rho(\tau) - \rho(\tau') \big|(\F) < \infty \qquad \text{and} \qquad
\big( \rho(\tau) - \rho(\tau') \big) (\F) = 0 \]
(where~$|.|$ denotes the total variation of a measure; see~\cite[\S28]{halmosmt}).
But the existence of minimizers has not yet been proven.
Nevertheless, the EL equations are well-defined in the following sense:
\begin{Def} \label{deflocmin}
Let~$(\rho, \H, \F)$ be a causal fermion system (possibly with~$\dim \H=\infty$
and~$\rho(\F)=\infty$). The measure~$\rho$ is a {\bf{local minimizer}} of the causal action principle
if the integral in~\eqref{Phi2def} is finite for all~$x \in \F$ and if the EL
equations~\eqref{Phiminimala} and~\eqref{Phiminimalb} hold for a suitable parameter~$\lambda \in \R$.
\end{Def} \noindent
Such local minimizers arise naturally when analyzing the continuum limit of
causal fermion systems (see~\cite{cfs}). Also, the physical examples 
in Sections~\ref{secexcurrent} and~\ref{secexEM} will be formulated for local minimizers in the
infinite-dimensional setting. Finally, the above notion of local minimizers is
of relevance in view of future extensions of the existence  theory to the infinite-dimensional setting.

Let~$\rho$ be a local minimizer of the causal action principle.
We again define {\em{space-time}} by~$M=\supp \rho$; it is a closed but in
general non-compact subset of~$\F\subset \Lin(\H)$.
We again define the function~$\ell$ by
\beq \label{elldef}
\ell(x) = \int_M \L_\kappa(x,y) \,d\rho(y)
\eeq
and for notational convenience set~$\nu = \lambda/2$.
By assumption, this function is well-defined and finite for all~$x \in \F$.
Moreover, the EL equations~\eqref{Phiminimala} and~\eqref{Phiminimalb} imply that
\beq \label{ELgen}
\begin{split}
\ell&(x) - \nu \,\tr(x) \qquad \text{is constant on~$M$} \\
\ell(x) - \nu \,\tr(x) &= \inf_{y \in U(x)} \big( \ell(y) - \nu \,\tr(y) \big) \qquad \text{for all~$x \in M$}
\end{split}
\eeq
(where~$U(x) \subset \F$ is again a neighborhood of~$x$).
However, the function~$\ell$ need not be integrable. In particular, the action~\eqref{Sdef}
may be infinite.

These EL equations imply analogs of the relations~\eqref{hyper} and~\eqref{Phi2def}.
Namely, evaluating the identity
\[ \frac{d}{dt} \big( \ell(tx) - \nu\, \tr(tx) \big) \big|_{t=1} = 0 \]
and using that the Lagrangian~\eqref{Lagrange} is homogeneous of degree two, one finds that on~$M$,
\[ 2 \ell(x) - \nu\, \tr(x) = 0 \:. \]
Combining this relation with~\eqref{ELgen}, one concludes that
on~$M$, the two terms in~\eqref{ELgen} are separately constant, i.e.
\beq \label{seprel}
\ell(x)  = -\inf_{y \in \F} \big( \ell(y) - \nu \,\tr(y) \big) = \frac{\nu}{2}\: \tr(x) \qquad \text{for all~$x \in M$}\:.
\eeq
These identities are very useful because they show that on~$M$,
both summands in~\eqref{ELgen} are separately constant. Moreover,
these relations make it possible to compute the Lagrange multiplier~$\nu$.

\subsection{Noether-Like Theorems} \label{seccfsnoether}
Let~$(\H, \F, \rho)$ be a causal fermion system, where~$\rho$ is a local minimizer
of the causal action (see Definition~\ref{deflocmin}).
We do not want to assume that~$\H$ is finite-dimensional nor that the total volume of~$\rho$
is finite. But we shall assume that~$\rho$ is {\bf{locally finite}} in the sense that~$\rho(K)< \infty$
for every compact subset~$K \subset \F$.

We again consider variations~$\Phi_\tau$ of~$M$ in~$\F$ described by a mapping~$\Phi$
of the form~\eqref{Phidef},
\beq \label{Phidef2}
\Phi : (-\tau_{\max}, \tau_{\max}) \times M \rightarrow \F
\qquad \text{with} \qquad \Phi(0,.) = \1 \:.
\eeq
Similar to Definition~\ref{defsymm}, the regularity of
the variation is defined by composing~$\Phi$ with an operator mapping to the real numbers.
However, we now compose both with~$\ell$ and with the trace operation.

\begin{Def} \label{defsymm2} A variation~$\Phi_\tau$ of the form~\eqref{Phidef2} is
is {\bf{continuous}} if the compositions
\[ \ell \circ \Phi ,\;\tr \circ \Phi : (-\tau_{\max}, \tau_{\max}) \times M \rightarrow \R \]
are continuous. If in addition their partial derivative~$\partial_\tau(\ell \circ \Phi)$
and~$\partial_\tau(\tr \circ \Phi)$ exist and are continuous on~$(-\tau_{\max}, \tau_{\max})
\times M \rightarrow \R$, then the variation is said to be {\bf{continuously differentiable}}.
\end{Def}

We now generalize Proposition~\ref{prpuseful} to the setting of causal fermion systems.
\begin{Prp} \label{prpuseful2} Let~$\Phi_\tau$ be a continuous variation of the form~\eqref{Phidef2}. Then
for any compact subset~$\Omega \subset M$,
\begin{align}
\int_M d\rho&(x) \int_\Omega d\rho(y) \Big( \L_\kappa\big( \Phi_\tau(x), y\big) - \L_\kappa(x,y) \Big) \label{id1n} \\
=\:& \int_\Omega \Big( \ell \big( \Phi_\tau(x)\big) - \ell(x) \Big)\: d\rho(x) \label{id2n} \\
&- \int_\Omega d\rho(x) \int_{M \setminus \Omega} d\rho(y) \:\Big( \L_\kappa\big( \Phi_\tau(x), y\big) -
\L_\kappa\big( x, \Phi_\tau(y) \big) \Big) \:. \label{id3n}
\end{align}
\end{Prp}
\Proof The subtle point is that~$M$ is in general non-compact, so that some of the integrals may diverge.
Therefore, we need to carefully consider the different integrals one after each other:
From~\eqref{seprel} we know that the functions~$\ell$ and~$\tr(x)$ are both constant on~$M$.
Moreover, the functions~$\ell \circ \Phi$ and~$\tr \circ \Phi$ are continuous on~$(-\tau_{\max}, \tau_{\max})
\times M$. As a consequence, it follows that for any {\em{compact}} subset~$\Omega \subset M$
and any~$\delta<\tau_{\max}$, the restriction
\[ \ell \circ \Phi \big|_{[-\delta, \delta] \times \Omega} \::\: 
[-\delta, \delta] \times \Omega \rightarrow \R \]
is a bounded function. Using that the Lagrangian is non-negative, this implies that
for any~$\tau \in (-\delta, \delta)$, the double integrals of the form
\[ \int_\Omega d\rho(x) \int_U d\rho(y) \:\L_\kappa\big( \Phi_\tau(x), y \big) \]
are well-defined and finite for any measurable subset~$U \subset M$. Moreover, one
may exchange the orders of integration using Tonelli's theorem
(i.e.\ the version of Fubini's theorem for non-negative integrands).
In particular, we conclude that the following integrals in~\eqref{id1n} and~\eqref{id3n} are
well-defined and finite,
\[ \int_M d\rho(x) \int_\Omega d\rho(y) \:\L_\kappa(x,y) \qquad \text{and} \qquad
\int_\Omega d\rho(x) \int_{M \setminus \Omega} d\rho(y) \:\L_\kappa\big( \Phi_\tau(x), y\big) \:. \]
For the integral in~\eqref{id2n}, we can argue similarly: We saw above that
the functions~$\ell \circ \Phi$ and~$\tr \circ \Phi$ are bounded on~$\{0\} \times M$ and continuous
on~$(-\tau_{\max}, \tau_{\max}) \times M$. Therefore, they are bounded on~$[-\delta, \delta] \times \Omega$,
implying that the integral in~\eqref{id2n} is well-defined and finite.

It remains to consider the two integrals
\beq \label{intdiverge}
\int_M d\rho(x) \int_\Omega d\rho(y) \:\L_\kappa\big( \Phi_\tau(x), y\big) \quad \text{and} \quad
\int_\Omega d\rho(x) \int_{M \setminus \Omega} d\rho(y) \:\L_\kappa\big( x, \Phi_\tau(y) \big) \:.
\eeq
These integrals could diverge. But since the integrand is non-negative, Tonelli's theorem
nevertheless allows us to exchange the two integrals.
Then the integrands of the two integrals coincide. The integration ranges
coincide up to the compact set~$\Omega \times \Omega$.
Therefore, the first integral in~\eqref{intdiverge} diverges if and only if the second integral diverges.
If this is the case, the left and the right side of the equation~\eqref{id1n}--\eqref{id3n}
both take the value~$+\infty$, so that the statement of the proposition holds.
In the case that the integrals in~\eqref{intdiverge} are both finite,
we can repeat the computation in the proof of Proposition~\ref{prpuseful}
and apply~\eqref{elldef} to obtain the result.
\QED

\begin{Def} \label{defgis2}
The variation~$\Phi_\tau$ is a {\bf{generalized integrated symmetry}} in
the space-time region~$\Omega \subset M$ if the following two identities hold:
\begin{gather}
\int_M d\rho(x) \int_\Omega d\rho(y) \Big( \L_\kappa\big( \Phi_\tau(x), y\big) - \L_\kappa(x,y) \Big) = 0 \label{Lagint} \\
\int_\Omega \Big( \tr \big( \Phi_\tau(x)\big) - \tr(x) \Big)\: d\rho(x) = 0 \:. \label{traceint}
\end{gather}
\end{Def}

Combining this definition with Proposition~\ref{prpuseful2} 
and the EL equations~\eqref{ELgen} immediately gives the following result:
\begin{Thm} \label{thmsymmgis2}
Let~$\rho$ be a local minimizer of the causal action (see Definition~\ref{deflocmin})
and $\Phi_\tau$ a continuously differentiable generalized integrated symmetry (see Definitions~\ref{defsymm2}
and~\ref{defgis2}). Then for any compact subset~$\Omega \subset M$, the following
surface layer integral vanishes,
\beq \label{conserve5}
\frac{d}{d\tau} \int_\Omega d\rho(x) \int_{M \setminus \Omega} d\rho(y) \:\Big( \L_\kappa\big( \Phi_\tau(x), y\big) -
\L_\kappa\big( x, \Phi_\tau(y) \big) \Big) \Big|_{\tau=0} = 0 \:.
\eeq
\end{Thm} \noindent
In order to explain the necessity of the condition~\eqref{traceint},
we point out that, although the functions~$\ell(x)$ and~$\tr(x)$
are both constant on~$M$ (see~\eqref{seprel}), this does not imply that transversal derivatives of
these functions vanish. Only for their specific linear combination in~\eqref{seprel} the
derivative vanishes on~$M$.
We also note that the condition for the trace~\eqref{traceint}, which did not appear in the compact setting,
can always be satisfied by rescaling the variation according to
\[ \Phi_\tau(x) \rightarrow \Phi_\tau(x)\:\frac{\tr(x)}{\tr\big( \Phi_\tau(x) \big)} \]
(note that by continuity, the trace in the denominator is non-zero for sufficiently small~$\tau$).
However, when doing so, the remaining condition~\eqref{Lagint} as well as the regularity conditions
of Definition~\ref{defsymm2} might become more involved.
This is the reason why we prefer to write two separate conditions~\eqref{Lagint} and~\eqref{traceint}.

The above results give rise to corollaries
which extend Theorems~\ref{thmsymmlag} and~\ref{thmsymmum} to the setting of
causal fermion systems.
\begin{Def} \label{defsymmnc} A variation~$\Phi_\tau$ of the form~\eqref{Phidef2}
is a {\bf{symmetry of the Lagrangian}} if
\beq \label{symmLagnc}
\L_\kappa \big( x, \Phi_\tau(y) \big) = \L_\kappa \big( \Phi_{-\tau}(x), y \big) \qquad
\text{for all~$\tau \in (-\tau_{\max}, \tau_{\max})$ and all~$x,y \in M$}\:.
\eeq
It is a {\bf{symmetry of the universal measure}} if
\[ (\Phi_\tau)_* \rho = \rho \qquad
\text{for all~$\tau \in (-\tau_{\max}, \tau_{\max})$}\:. \]
Moreover, it {\bf{preserves the trace}} if
\[ \tr \big( \Phi_\tau(x)\big) = \tr(x) \qquad \text{for all~$\tau \in (-\tau_{\max}, \tau_{\max})$
and all~$x \in M$}\:. \]
\end{Def}

\begin{Corollary} \label{corsymmlag}
Let~$\rho$ be a local minimizer of the causal action (see Definition~\ref{deflocmin})
and $\Phi_\tau$ a continuously differentiable variation. Assume that~$\Phi_\tau$ is a
symmetry of the Lagrangian and preserves the trace.
Then for any compact subset~$\Omega \subset M$, the conservation law~\eqref{conserve5} holds.
\end{Corollary}

\begin{Corollary} \label{corsymmum}
Let~$\rho$ be a local minimizer of the causal action (see Definition~\ref{deflocmin})
and $\Phi_\tau$ a continuously differentiable variation. Assume that~$\Phi_\tau$ 
is a symmetry of the universal measure and preserves the trace.
Then for any compact subset~$\Omega \subset M$, the conservation law~\eqref{conserve5} holds.
\end{Corollary} \noindent
These corollaries follow immediately by calculations similar to~\eqref{calc1} and~\eqref{calc2}.

\section{Example: Current Conservation} \label{secexcurrent}
This section is devoted to the important example of current conservation,
also referred to as charge conservation.
For Dirac particles, the electric charge is (up to a multiplicative constant)
given as the integral over the probability density. Therefore, charge conservation
also corresponds to the conservation of the probability integral in quantum mechanics.
In the context of the classical Noether theorem, charge conservation is a consequence
of an internal symmetry of the system, which can be described by a phase transformation~\eqref{globalphase}
of the wave function and is often referred to as global gauge symmetry.
As we shall see in Section~\ref{secgenchargecons}, causal fermion systems also have such
an internal symmetry, giving rise to a general class of conservation laws (see Theorem~\ref{thmcurrent}).
In Section~\ref{seccurcor}, these conservation laws are evaluated for Dirac spinors
in Minkowski space, giving a correspondence to the conservation of the Dirac current
(see Theorem~\ref{thmcurrentmink} and Corollary~\ref{corcurrent}).
In Section~\ref{secremark}, we conclude with a few clarifying remarks.

\subsection{A General Conservation Law} \label{secgenchargecons}
Let~$\scrA$ be a bounded symmetric operator on~$\H$ and
\beq \label{Utaudef}
\scrU_\tau := \exp(i \tau \scrA)
\eeq
be the corresponding one-parameter family of unitary transformations.
We introduce the mapping
\beq \label{varunit}
\Phi_\tau \,:\, \R \times \F \rightarrow \F\:,\qquad \Phi(\tau, x) = \scrU_\tau \,x\, \scrU_\tau^{-1} \:.
\eeq
Restricting this mapping to~$(-\tau_{\max}, \tau_{\max}) \times M$, we obtain
a variation~$(\Phi_\tau)_{\tau \in (-\tau_{\max}, \tau_{\max})}$ of the form~\eqref{Phidef2}.

\begin{Lemma} \label{PhiSymmLag} The variation~$\Phi_\tau$ given by~\eqref{varunit} is a symmetry of the
Lagrangian and preserves the trace (see Definition~\ref{defsymmnc}).
\end{Lemma}
\Proof Since~$\Phi_\tau(x)$ is unitarily equivalent to~$x$, they obviously have the same trace.
In order to prove~\eqref{symmLagnc}, we first recall that the Lagrangian~$\L_\kappa(x,y)$
is defined in terms of the spectrum of the operator product~$xy$ (see~\eqref{Lagrange}).
The calculation
\[ x \: \Phi_\tau(y) = x \; \scrU_\tau \,y\, \scrU_\tau^{-1}
= \scrU \,\big( \scrU_\tau^{-1} \,x \,\scrU_\tau \; y \big) \,\scrU_\tau^{-1}
= \scrU \,\big( \Phi_{-\tau}(x) \: y \big) \, \scrU_\tau^{-1} \]
shows that the operators~$x \,\Phi_\tau(y)$ and~$\Phi_{-\tau}(x) \,y$ are unitarily equivalent
and therefore isospectral. This concludes the proof.
\QED

It remains to verify whether the variation~$\Phi_\tau$ is continuously differentiable
in the sense of Definition~\ref{defsymm2}. For the trace, this is obvious because~$\Phi_\tau$
leaves the trace invariant, so that~$\tr \,\circ\, \Phi_\tau(\tau, x) = \tr(x)$, which clearly depends
continuously on~$x$ (in the topology induced by the $\sup$-norm~\eqref{supnorm}).
For~$\ell \circ \phi$, we cannot in general expect differentiability because the Lagrangian~$\L_\kappa$
is only Lipschitz continuous in general. Therefore, we must include the differentiability of~$\ell \circ \phi$
as an assumption in the follo\-wing theo\-rem.

\begin{Thm} \label{thmcurrent}
Given a bounded symmetric operator~$\scrA$ on~$\H$, we let~$\Phi_\tau$ be the
variation~\eqref{varunit}. Assume that the mapping~$\ell \circ \Phi \::\:
(-\tau_{\max}, \tau_{\max}) \times M \rightarrow \R$
is continuously differentiable in the sense that it is continuous and that~$\partial_\tau (\ell \circ \Phi)$
exists and is also continuous on~$(-\tau_{\max}, \tau_{\max}) \times M$.
Then for any compact subset~$\Omega \subset M$, the conservation law~\eqref{conserve5} holds.
\end{Thm}

\subsection{Correspondence to Dirac Current Conservation} \label{seccurcor}
The aim of this section is to relate the conservation law of Theorem~\ref{thmcurrent} to
the usual current conservation in relativistic quantum mechanics in Minkowski space.

To this end, we consider causal fermion systems~$(\F, \H, \rho^\varepsilon)$
describing the regularized Dirac sea vacuum in Minkowski space~$(\scrM, \la .,. \ra)$.
We briefly recall the construction (for the necessary preliminaries see~\cite[Section~2]{cfsrev}, \cite{cfs},
\cite[Section~4]{lqg} or the introductory paper~\cite{dice2014}).
As in~\cite[Chapter~3]{cfs} we consider three generations of Dirac particles
of masses~$m_1$, $m_2$ and~$m_3$ (corresponding to the three generations of elementary particles
in the standard model; three generations are necessary in order to obtain well-posed equations
in the continuum limit). Denoting the generations by an index~$\beta$, we consider the
Dirac equations
\beq \label{Direqns}
(i \Pdd - m_\beta) \,\psi_\beta = 0 \qquad (\beta=1,2,3)\:.
\eeq
On solutions~$\psi = (\psi_\beta)_{\beta=1,2,3}$, we consider the scalar product
\[ ( \psi | \phi) := 2 \pi \sum_{\beta=1}^3 \int_{\R^3} (\overline{\psi}_\beta \gamma^0 \phi_\beta)(t, \vec{x})\: d^3x \:. \]
The Dirac equation has solutions on the upper and lower mass shell, which have positive
respectively negative energy.
In order to avoid potential confusion with other notions of energy, we here prefer the notion of
solutions of positive and negative {\em{frequency}}.
We choose~$\H$ as the subspace spanned by all solutions of negative frequency,
together with the scalar product~$\la .|. \ra_\H := ( .|.)|_{\H \times \H}$.
We now introduce an ultraviolet regularization
(for details see~\cite[Section~2]{cfsrev}) and denote the regularized quantities by
a superscript~$\varepsilon$. Now the local correlation operators are defined by
\[ \la \psi^\varepsilon \,|\, F^\varepsilon(x)\, \phi^\varepsilon \ra_\H
= - \sum_{\alpha, \beta=1}^3 \overline{\psi_\alpha^\varepsilon(x)} \phi_\beta^\varepsilon(x)
\qquad \text{for all~$\psi,\phi \in \H$}\:. \]
Next, the universal measure is defined as the push-forward of the Lebesgue measure~$d\mu = d^4x$,
\[ \rho^\varepsilon := (F_\varepsilon)_*(\mu) \:. \]
Then~$(\H, \F, \rho^\varepsilon)$ is a causal fermion system of spin dimension two.
As shown in~\cite[Chapter~1]{cfs}, the kernel of the fermionic projector~$P(x,y)$
converges as~$\varepsilon \searrow 0$ to the distribution
\beq \label{Pvac}
P(x,y) = \sum_{\beta=1}^3 \int \frac{d^4k}{(2 \pi)^4}\: (\slashed{k}+m_\beta)\:
\delta \big(k^2-m_\beta^2 \big)\: e^{-ik(x-y)}
\eeq
(this configuration is also referred to as three generations in a single sector;
see~\cite[Chapter~3]{cfs}). We remark that our ansatz can be generalized by
introducing so-called weight factors (see~\cite{reg} and Remark~\ref{remweights} below).

\begin{figure}
\psscalebox{1.0 1.0} 
{
\begin{pspicture}(0,-0.9569027)(10.911616,0.9569027)
\definecolor{colour0}{rgb}{0.8,0.8,0.8}
\psframe[linecolor=colour0, linewidth=0.02, fillstyle=solid,fillcolor=colour0, dimen=outer](9.869394,0.56531954)(5.789394,-0.8213471)
\pspolygon[linecolor=colour0, linewidth=0.02, fillstyle=solid,fillcolor=colour0](0.033838496,-0.34579158)(0.19828294,-0.23023602)(0.35828295,-0.15023603)(0.5760607,-0.03912491)(0.94939405,0.10309731)(1.389394,0.21865287)(1.8160607,0.27198622)(2.1227274,0.28531954)(2.4338386,0.26754177)(2.7982829,0.17865287)(3.1227274,0.067541756)(3.4427273,-0.083569355)(3.6827273,-0.21690269)(3.598283,-0.33690268)(3.3938384,-0.46134713)(3.1316164,-0.5991249)(2.8560607,-0.7191249)(2.478283,-0.7946805)(2.1627274,-0.84356934)(1.8160607,-0.85690266)(1.5182829,-0.8480138)(1.0827274,-0.78134716)(0.62939405,-0.63912493)(0.2827274,-0.4791249)
\rput[bl](1.4738384,-0.41468045){$\Omega$}
\rput[bl](7.458283,-0.27690268){\normalsize{$\Omega$}}
\psline[linecolor=black, linewidth=0.04, arrowsize=0.09300000000000001cm 1.0,arrowlength=1.7,arrowinset=0.3]{->}(1.3138385,0.18531954)(1.2116163,0.6919862)
\rput[bl](1.4160607,0.45643064){$\nu$}
\rput[bl](10.027172,0.41865286){\normalsize{$t=t_1$}}
\rput[bl](10.031616,-0.9569027){\normalsize{$t=t_0$}}
\psbezier[linecolor=black, linewidth=0.04](0.011616275,-0.3569027)(0.704925,0.13707085)(1.4876974,0.28290415)(2.0582829,0.28754175)(2.6288686,0.29217938)(3.0587788,0.14209865)(3.7116163,-0.23690268)
\psbezier[linecolor=black, linewidth=0.04](0.014949608,-0.34134713)(0.5271472,-0.62181807)(0.9588085,-0.8170959)(1.6016163,-0.86023605)(2.244424,-0.90337616)(3.12989,-0.7812347)(3.7149496,-0.22134712)
\psline[linecolor=black, linewidth=0.04, arrowsize=0.09300000000000001cm 1.0,arrowlength=1.7,arrowinset=0.3]{->}(2.4071717,-0.81468046)(2.3049495,-0.3080138)
\psline[linecolor=black, linewidth=0.04](5.789394,0.57420844)(9.864949,0.57420844)
\psline[linecolor=black, linewidth=0.04](5.789394,-0.8213471)(9.869394,-0.8257916)
\psline[linecolor=black, linewidth=0.04, arrowsize=0.09300000000000001cm 1.0,arrowlength=1.7,arrowinset=0.3]{->}(6.669394,0.5764306)(6.6738386,1.069764)
\psline[linecolor=black, linewidth=0.04, arrowsize=0.09300000000000001cm 1.0,arrowlength=1.7,arrowinset=0.3]{->}(8.660505,-0.82356936)(8.664949,-0.33023602)
\rput[bl](2.5138385,-0.60579157){$\nu$}
\rput[bl](8.864949,-0.6413471){$\nu$}
\rput[bl](6.8649497,0.7630973){$\nu$}
\end{pspicture}
}
\caption{Choice of the space-time region~$\Omega \subset \scrM$.}
\label{fignoether2}
\end{figure}
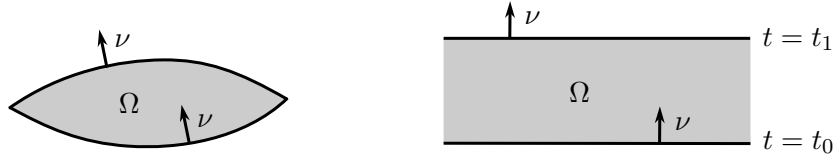

We want to apply Theorem~\ref{thmcurrent}.
Since in this theorem, the set~$\Omega$ must be compact,
we choose it as a lens-shaped region whose boundary is composed of two
space-like hypersurfaces (see the left of Figure~\ref{fignoether2}).
Considering a sequence
of compact sets~$\Omega_n$ which exhaust the region~$\Omega$
between two Cauchy surfaces at times~$t=t_0$
and~$t=t_1$, the surface layer integral~\eqref{conserve5} reduces to the difference
of surface layers integrals at times~$t \approx t_0$ and~$t \approx t_1$.
The detailed analysis (which will be carried out below) gives the following result:
\begin{Thm} {\bf{(current conservation)}} \label{thmcurrentmink}
Let~$(\H, \F, \rho^\varepsilon)$ be local minimizers of the causal action which
describe the Minkowski vacuum~\eqref{Pvac}.
Considering the limiting procedure explained in Figure~\ref{fignoether2} and taking the
continuum limit, the conservation laws of Theorem~\ref{thmcurrent} go over to
a linear combination of the probability integrals in every generation.
More precisely, there are non-negative constants~$c_\beta$ such that
for all~$u \in \H$ for which~$\psi^u$ is a negative-frequency solution of the Dirac equation,
the surface layer integral~\eqref{conserve5} goes over the equation
\beq \label{thmcurr1}
\sum_{\beta=1}^3 m_\beta\, c_\beta \int_{t=t_0} \Sl \psi_\beta^u(x) | \gamma^0 \psi_\beta^u(x) \Sr\:d^3x
= \sum_{\beta=1}^3 m_\beta\, c_\beta \int_{t=t_1} \Sl \psi_\beta^u(x) | \gamma^0 \psi_\beta^u(x) \Sr\:d^3x \:.
\eeq
\end{Thm} \noindent
The constants~$c_\beta$ depend on properties of the distribution~$\hat{Q}$ in the continuum limit,
as will be specified in Definition~\ref{def611} and~\eqref{USymm20} below.

Before coming to the proof, we explain the statement and significance of this theorem.
We first note that the restriction to negative-frequency solutions is needed
because the description of positive-frequency solutions involves the 
so-called mechanism of microscopic mixing which for brevity we cannot address in this
paper (see however Remark~\ref{remmicro} below).
Next, we point out that the theorem implies the statement that the function~$\ell \circ \Phi$ in
Theorem~\ref{thmcurrent} is continuously differentiable in the continuum limit.
However, this does not necessarily mean that this differentiability statement
holds for any local minimizer~$(\H, \F, \rho^\varepsilon)$ with regularization.
This rather delicate technical point will be discussed in Remark~\ref{remdiscuss} below.

Considering Cauchy hyperplanes in~\eqref{thmcurr1} is indeed no
restriction because the theorem can be extended immediately to
general Cauchy surfaces:
\begin{Corollary} {\bf{(current conservation on Cauchy surfaces)}}  \label{corcurrent}
Let~$\scrN_0, \scrN_1$ be two Cauchy surfaces in Minkowski space, where~$\scrN_1$
lies to the future of~$\scrN_0$.
Then, under the assumptions of Theorem~\ref{thmcurrentmink},
the conservation law of Theorem~\ref{thmcurrent} goes over to the conservation law
for the current integrals
\beq \label{consCauchy}
\sum_{\beta=1}^3 m_\beta\, c_\beta \int_{\scrN_0} \Sl \psi_\beta^u | \slashed{\nu} \psi_\beta^u \Sr\:d\mu_{\scrN_0}
= \sum_{\beta=1}^3 m_\beta\, c_\beta \int_{\scrN_1} \Sl \psi_\beta^u | \slashed{\nu} \psi_\beta^u \Sr\:d\mu_{\scrN_1} \:,
\eeq
where~$\nu$ denotes the future-directed normal.
\end{Corollary}
\Proof We choose~$\Omega$ as the space-time region between the two Cauchy surfaces.
Using that the integrand in~\eqref{conserve5} is anti-symmetric in its arguments~$x$ and~$y$,
the integration range can be rewritten as
\begin{align}
\int_\Omega &d\rho(x) \int_{M \setminus \Omega} d\rho(y) \:\Big( \L_\kappa\big( \Phi_\tau(x), y\big) -
\L_\kappa\big( x, \Phi_\tau(y) \big) \Big) \nonumber \\
&= \int_{J^\wedge(\scrN_1)} d\rho(x) \int_{J^\vee(\scrN_1)} d\rho(y)
\:\Big( \L_\kappa\big( \Phi_\tau(x), y\big) - \L_\kappa\big( x, \Phi_\tau(y) \big) \Big) \label{sl1} \\
&\quad\: -\int_{J^\wedge(\scrN_0)} d\rho(x) \int_{J^\vee(\scrN_0)} d\rho(y)
\:\Big( \L_\kappa\big( \Phi_\tau(x), y\big) - \L_\kappa\big( x, \Phi_\tau(y) \big) \Big) \:, \label{sl2}
\end{align}
where~$J^\wedge$ and~$J^\vee$ denote the causal past and causal future, respectively.
For ease in notation, we refer to the integrals in~\eqref{sl1} as a
{\em{surface layer integral over}}~$\scrN_1$.
Thus the surface layer integral in~\eqref{conserve5} is the difference
of two surface layer integrals over the Cauchy surfaces~$\scrN_0$ and~$\scrN_1$.

In order to compute for example the surface layer integral over~$\scrN_0$,
one chooses~$\Omega$ as the region between the
Cauchy surface~$\scrN_0$ and the Cauchy surface~$t=t_0$
(for sufficiently small~$t_0$; in case that these Cauchy surfaces intersect for every~$t_0$, one
modifies~$\scrN_0$ near the asymptotic end without affecting our results).
Applying the conservation law of Theorem~\ref{thmcurrent}
to this new region~$\Omega$, one concludes that the the surface layer integral over~$\scrN_0$
coincides with the surface layer integral at time~$t \approx t_0$. The latter surface layer
integral, on the other hand, was computed in Theorem~\ref{thmcurrentmink}
to go over to the sum of the probability integrals in~\eqref{thmcurr1}.
Finally, the usual current conservation for the Dirac dynamics shows that the
the integrals in~\eqref{thmcurr1} coincide with the surface integral over~$\scrN_0$
in~\eqref{consCauchy}. This concludes the proof.
\QED
Using similar arguments, Theorem~\ref{thmcurrentmink} can also be extended
to interacting systems (see Remark~\ref{reminteract} below). \\[-0.7em]

The remainder of this section is devoted to the proof of Theorem~\ref{thmcurrentmink}.
We first rewrite the causal action principle in terms of the kernel of the fermionic projector
(for details see~\cite[\S1.1]{cfs}). The kernel of the fermionic projector~$P(x,y)$ is defined by
\beq \label{Pxydef}
P(x,y) = \pi_x \,y|_{S_y} \::\: S_y \rightarrow S_x \:.
\eeq
The {\em{closed chain}} is defined as the product
\[ A_{xy} = P(x,y)\, P(y,x) \::\: S_x \rightarrow S_x\:. \]
The nontrivial eigenvalues~$\lambda^{xy}_1, \ldots, \lambda^{xy}$ of the operator~$xy$
coincide with the eigenvalues of the closed chain.
Moreover, it is useful to express~$P(x,y)$ in terms of the wave evaluation operator defined by
\beq
\label{weo}
\Psi(x) \::\: \H \rightarrow S_x\:, \qquad u \mapsto \psi^u(x) = \pi_x u \:. 
\eeq
Namely,
\[ x = - \Psi(x)^* \,\Psi(x)  \qquad \text{and} \qquad P(x,y) = -\Psi(x)\, \Psi(y)^*\:. \]

Our task is to compute the term~$\L_\kappa(\Phi_\tau(x), y)$ in~\eqref{conserve5} for~$x, y \in M$.
The detailed computations in ~\cite[\S3.6.1]{cfs} show that the
fermionic projector of the Minkowski vacuum satisfies the
EL equations in the continuum limit for~$\kappa=0$
(in our setting, this result means that the measures~$\rho^\varepsilon$
are local minimizers in the sense of Definition~\ref{deflocmin} in the limiting case~$\varepsilon \searrow 0$).
Therefore, we may set~$\kappa$ to zero.
Thus our task is to compute the term~$\L(\Phi_\tau(x), y)$.
In preparation, we compute~$P(\Phi_\tau(x), y)$.
To this end, we first note that
\beq \label{isospec}
\begin{split}
\Phi_\tau(x)\, y &= \scrU_\tau \,x\, \scrU_\tau^{-1}\; y
= \scrU_\tau \,\Psi(x)^* \,\Psi(x)\, \scrU_\tau^{-1}\; \Psi(y)^* \,\Psi(y) \\
&\simeq \Psi(x)\, \scrU_\tau^{-1}\; \Psi(y)^* \,\Psi(y)\, \scrU_\tau \,\Psi(x)^* \:,
\end{split}
\eeq
where in the last line we cyclically commuted the operators
and $\simeq$ means that the operators are isospectral (up to irrelevant zeros in the spectrum).
Therefore, introducing the notations
\begin{gather}
\Psi_\tau(x) = \Psi(x)\, \scrU_\tau^{-1} \::\: \H \rightarrow S_x \\
P\big( \Phi_\tau(x), y \big) = -\Psi_\tau(x)\, \Psi(y)^* \:, \qquad
P\big( y, \Phi_\tau(x) \big) = -\Psi(y)\, \Psi_\tau(x)^* \label{Pmodform}
\end{gather}
one sees that the operator product~$\Phi_\tau(x)\, y$ is isospectral to the modified closed chain
\beq \label{chainform}
P\big( \Phi_\tau(x), y \big)\: P\big( y, \Phi_\tau(x) \big) \:.
\eeq
Considering the Lagrangian as a function of this modified closed chain, the variation
is described in a form suitable for computations.

For clarity, we explain in which sense the kernel of the fermionic projector as given by~\eqref{Pmodform}
agrees with the abstract definition~\eqref{Pxydef},
\beq \label{Ptauabstract}
P\big( \Phi_\tau(x), y \big) = \pi_{\Phi_\tau(x)} y \:.
\eeq
It is a subtle point that the point~$\Phi_\tau(x) \in \F$ depends on~$\tau$, so that space-time
itself changes. However, when identifying the spin space~$S_{\Phi_\tau(x)}$ with a corresponding
spinor space in Minkowski space, the base point~$x \in \scrM$ should be kept fixed.
Therefore, the spin space~$S_{\Phi_\tau(x)}$ is to be identified with the spinor space~$S_x \scrM$.
For each~$\tau$, this can be accomplished as explained above.
This identification made, the kernel~\eqref{Pmodform} indeed agrees with~\eqref{Ptauabstract}.
The reason why we do not give the details of this construction is that
the computation~\eqref{isospec} already shows that the Lagrangian
can be computed with the closed chain~\eqref{chainform}, and this is all we need for what follows.

We now choose~$\scrA =\pi_{\la u \ra}$ as the projection on the one-dimensional subspace
generated by a vector~$u \in \H$ and let~$\pi_{\la u \ra^\perp}$ be the projection on
the orthogonal complement of~$u$. Then
\begin{align*}
\Psi_\tau(x) &= \Psi(x) \; \big( \pi_{\la u \ra^\perp} + e^{-i \tau}\, \pi_{\la u \ra} \big) \\
P\big(\Phi_\tau(x),y \big) &= -\Psi(x) \; \big( \pi_{\la u \ra^\perp} + e^{-i \tau}\, \pi_{\la u \ra} \big)\: \Psi(y)^* \\
&= P(x,y) + (1- e^{-i \tau}) \;\Psi(x) \:\pi_{\la u \ra}\: \Psi(y)^* \:.
\end{align*}
Normalizing~$u$ such that~$\la u|u \ra_\H=1$, the last equation can be written
in the form that for any~$\chi \in S_y$,
\[ P\big(\Phi_\tau(x),y \big) \, \chi
= P(x,y)\, \chi + (1- e^{-i \tau}) \;\psi^u(x)\; \Sl \psi^u(y) \,|\, \chi \Sr_y \:. \]

We now compute the first order variation.
\beq \label{delP} \begin{split}
\frac{d}{d\tau} P\big(\Phi_\tau(x),y \big) \big|_{\tau=0} \:\chi &=
i \psi^u(x)\; \Sl \psi^u(y)\,|\, \chi \Sr_y =: \delta P(x,y)\, \chi \\
\frac{d}{d\tau} P\big(y, \Phi_\tau(x)\big) \big|_{\tau=0} &= \big(\delta P(x,y) \big)^*
\end{split}
\eeq
The variation of the Lagrangian can be written as (cf.~\cite[Section~5.2]{PFP}
or~\cite[Section~1.4]{cfs})
\begin{align} \label{LTrQ} \begin{split}
\delta \L(x,y) \;&\!:=  \frac{d}{d\tau}\: \L\big( \Phi_\tau(x),y \big)\big|_{\tau=0} \\
&= \Tr_{S_y} \big( Q(y,x)\, \delta P(x,y) \big) + \Tr_{S_x} \big( Q(x,y)\, \delta P(x,y)^* \big) \\
&= i \, \Sl \psi^u(y) \,|\, Q(y,x) \,\psi^u(x) \Sr_y
-i \,\Sl \psi^u(x) \,|\, Q(x,y) \,\psi^u(y) \,\Sr_x \:, \end{split}
\end{align}
where in the last line we used~\eqref{delP}, and~$Q(x,y)$ is a distributional kernel
to be specified below.
Using that the kernel~$Q(x,y)$ is symmetric in the sense that
\[ Q(x,y)^* = Q(y,x)\:, \]
we can write the variation of the Lagrangian in the compact form
\[ \delta \L(x,y) = -2 \im \big( \Sl \psi^u(y) \,|\, Q(y,x) \,\psi^u(x) \Sr_y \big) \:. \]
Using this identity, the surface layer integral in~\eqref{conserve5} can be written as
\[ \int_\Omega d^4x \int_{\scrM \setminus \Omega} d^4y \;\im \big( \Sl \psi^u(y) \,|\, Q(y,x) \,\psi^u(x) \Sr_y \big)
= 0 \:. \]

Taking the liming procedure as shown in Figure~\ref{fignoether2}, it suffices to
consider a surface layer integral at a fixed time~$t_0$,
which for convenience we choose equal to zero. Thus our task is to compute the double integral
\beq \label{inttask}
J := \int_{t \geq 0} d^4x \int_{t < 0} d^4y \:\im \big( \Sl \psi^u(y) \,|\, Q(y,x) \,\psi^u(x) \Sr \big) \:.
\eeq
Here we omitted the subscript~$y$ at the spin scalar product because in Minkowski space
all spinor spaces can be naturally identified.

In order to explain our method for computing the integrals in~\eqref{inttask},
we first state a simple lemma where integrals of this type are computed.
As will be explained below, this lemma cannot be applied to our problem
for technical reasons, but it nevertheless clarifies the structure of our results.
\begin{Lemma} \label{lemmagen}
Let~$f \::\: \scrM \times \scrM \rightarrow \R$ be an integrable function
with the following properties:
\begin{itemize}
\item[(a)] $f$ is anti-symmetric, i.e.\ $f(x,y) = -f(y,x)$.
\item[(b)] $f$ is homogeneous in the sense that it depends only on the difference vector~$y-x$.
\item[(c)] The following integral is finite,
\beq \label{finint}
\int_{\scrM} \big|x^0\: f(x,0) \big|\: d^4x < \infty \:.
\eeq
\end{itemize}
Then
\beq \label{dint}
\int_{-\infty}^0 dt \int_0^\infty dt' \int_{\R^3} d^3y \:f \big( (t,\vec{x}), (t', \vec{y}) \big) = \frac{i}{2}\:
\frac{\partial}{\partial k^0} \hat{f}(k) \Big|_{k=0} \:,
\eeq
where~$\hat{f}$ is the Fourier transform, i.e.\
\beq \label{fFourier}
f(x,y) = \int \frac{d^4k}{(2 \pi)^4} \:\hat{f}(k)\: e^{-ik (x-y)} \:.
\eeq
\end{Lemma}
\Proof Substituting~\eqref{fFourier} into the left side of~\eqref{dint}, we can carry out the spatial integral to obtain
\beq \label{doubleint}
\int_{-\infty}^0 dt \int_0^\infty dt' \int_{\R^3} d^3y \:f \big( (t,\vec{x}), (t', \vec{y}) \big) =
\int_{-\infty}^0 dt \int_0^\infty dt' \,g(t-t') \:,
\eeq
where
\beq \label{gFourier}
g(\tau) =  \int_{\R^3} f\big((\tau, \vec{x}), (0, \vec{y}) \big)\: d^3y
= \int_{-\infty}^\infty \frac{d\omega}{2 \pi} \:\hat{f}\big( (\omega, \vec{0} ) \big)\: e^{-i\omega \tau} \:.
\eeq
We now transform variables in the inner integral in~\eqref{doubleint},
\[ \int_0^\infty g(t-t')\: dt' = \int_{-\infty}^t g(\tau)\: d\tau = \int_{-\infty}^0 g(\tau)\: \Theta(t-\tau) \: d\tau \:. \]
Using~\eqref{finint} and~\eqref{gFourier}, we know that
\[ \iint_{\R^- \times \R^-} \big| g(\tau)\: \Theta(t-\tau) \big| \: dt\, d\tau
= \int_{-\infty}^0 \big| \tau\: g(\tau)\big| \: d\tau
\leq \int_{\scrM} \big|x^0\: f(x,0) \big|\: d^4x < \infty \:. \]
Hence in~\eqref{doubleint} we may switch the order of integration according to Fubini's theorem to obtain
\begin{align*}
\int_{-\infty}^0 &dt \int_0^\infty dt' \,g(t-t')
= \int_{-\infty}^0 d\tau \,g(\tau) \int_{-\infty}^0 dt  \: \Theta(t-\tau) \\
&= \int_{-\infty}^0 d\tau \,g(\tau) \int_\tau^0 dt  = - \int_{-\infty}^0 d\tau \,\tau\, g(\tau) 
=  -\frac{1}{2} \int_{-\infty}^\infty d\tau \,\tau\, g(\tau) \:,
\end{align*}
where in the last step we used the anti-symmetry of~$g$.
Now  we insert~\eqref{gFourier} and apply Plancherel's theorem,
\begin{align*}
\int_{-\infty}^\infty &dt \int_0^\infty dt' \,g(t-t')
=-\frac{i}{2} \int_{-\infty}^\infty d\tau \int_{-\infty}^\infty \frac{d\omega}{2 \pi} \:\hat{f}\big( (\omega, \vec{0}) \big)\:
\frac{\partial}{\partial \omega} e^{-i\omega \tau} \\
&=\frac{i}{2} \int_{-\infty}^\infty d\tau \int_{-\infty}^\infty \frac{d\omega}{2 \pi} \:
\Big(\frac{\partial}{\partial \omega} \hat{f}\big( (\omega, \vec{0}) \big) \Big)\:
e^{-i\omega \tau} = \frac{i}{2} \:\frac{\partial}{\partial \omega} \hat{f}\big( (\omega, \vec{0}) \big) \Big|_{\omega=0}\:.
\end{align*}
This concludes the proof.
\QED
In order to apply this lemma to our problem, we would have to show that the integrand
in~\eqref{inttask} satisfies the condition~\eqref{finint}. As we shall now explain,
this condition will indeed {\em{not}} be satisfied, making it necessary to modify the method.

Let us specify the kernel~$Q(x,y)$. To this end, we make use of the fact
that the fermionic projector of the vacuum should correspond to a stable
minimizer of the causal action. This is made mathematically precise
in the so-called state stability analysis carried out in~\cite[Section~5.6]{PFP},
\cite{vacstab} and~\cite{reg}. The detailed analysis of the continuum
limit in~\cite[Chapter~3]{cfs} shows that in order to obtain well-defined field equations
in the continuum limit, the number of generations must be equal to three.
Therefore, we now consider an unregularized fermionic projector of the vacuum
involving a sum of three Dirac seas~\eqref{Pvac}.
The corresponding kernel~$Q(x,y)$ obtained in the continuum limit depends only on
the difference vector~$y-x$ and can thus be written as the Fourier transform
of a distribution~$\hat{Q}(k)$,
\[ Q(x,y) = \int \frac{d^4k}{(2 \pi)^4} \:\hat{Q}(k)\: e^{-ik (x-y)} \:. \]
The state stability analysis in~\cite[Section~5.6]{PFP} implies that the Fourier
transform~$\hat{Q}$ has the form as specified in the next definition
(cf.~\cite[Definition~5.6.2]{PFP}).

\begin{Def} \label{def611}\index{state stability}
The fermionic projector of the vacuum~\eqref{Pvac} is called {\bf{state stable}}
if the corresponding operator $\hat{Q}(k)$ is well-defined inside the
lower mass cone
\[ \mathcal{C}^\land := \{ k \in \R^4 \,|\, k^i k_i >0 \text{ and } k^0<0 \} \]
and can be written as
\beq \label{62f}
\hat{Q} (k) = a\:\frac{k\slsh}{|k|} + b
\eeq
with continuous real functions $a$ and $b$ on $\mathcal{C}^\land$ having
the following properties:
\begin{itemize}
\item[(i)] $a$ and $b$ are Lorentz invariant,
\[ a = a(k^2)\:,\qquad b = b(k^2) \:. \]
\item[(ii)] $a$ is non-negative.
\item[(iii)] The function $a+b$ is minimal on the mass shells,
\beq \label{abmin}
(a+b)(m^2_\beta) = \inf_{q \in {\mathcal{C}}^\land} (a+b)(q^2) \quad\mbox{for~$\beta=1,2,3$}\:.
\eeq
\end{itemize}
\end{Def}
We point out that, according to this definition, the function~$\hat{Q}(k)$ does {\em{not}} need to be {\em{smooth}},
but only continuous. In particular, Lemma~\ref{lemmagen} cannot be
applied, because the derivative in~\eqref{dint} is ill-defined.
If~$\hat{Q}(k)$ were smooth, its Fourier transform~$Q(x,y)$ would decay rapidly as~$(y-x)^2 \rightarrow \pm \infty$.
In this case, $Q(x,y)$ would be of short range as explained in Section~\ref{secsli},
except that~\eqref{shortrange} would have to be replaced by
the statement that~$\L(x,y)$ is very small if~$|(y-x)^2|>\delta$
(and~$\L(x,y)$ could indeed be made arbitrarily small by increasing~$\delta$).
The fact that~$\hat{Q}(k)$ does not need to be differentiable implies that~$Q(x,y)$
does not need to decay rapidly, also implying that the condition~\eqref{finint} may be violated.

In fact, this non-smoothness in momentum space
will be of importance in the following computation. Moreover, our results will depend
only on the behavior~$\hat{Q}(k)$ in a neighborhood of the mass shells~$k^2=m_\beta^2$.
Therefore, the crucial role will be played by the regularity of~$\hat{Q}$ on the mass shells.
In order to keep the setting as simple as possible, we shall assume that the functions~$a$
and~$b$ in~\eqref{62f} are {\em{semi-differentiable}} on the mass shells, meaning that the
left and right derivatives exist. For the resulting semi-derivatives of~$\hat{Q}$ we use the notation
\beq \label{semidiff}
\begin{split}
\partial_\omega^+ \hat Q( -\omega_{\beta, \vec k}, \vec k) &= \lim_{h \searrow 0}\:
\frac{1}{h}\, \Big( \hat Q( -\omega_{\beta, \vec k}+h, \vec k) - \hat Q( -\omega_{\beta, \vec k}, \vec k) \Big)  \\
\partial_\omega^- \hat Q( -\omega_{\beta, \vec k}, \vec k) &= \lim_{h \nearrow 0}\:
\frac{1}{h}\, \Big( \hat Q( -\omega_{\beta, \vec k}+h, \vec k) - \hat Q( -\omega_{\beta, \vec k}, \vec k) \Big) \:,
\end{split} \eeq
where~$\omega_{\beta, \vec k}$ is given by the dispersion relation
\beq \label{disperse}
\omega_{\beta, \vec k} = \sqrt{m_\beta^2 + |\vec{k}|^2}\:.
\eeq
The parameters~$c_\beta$ in Theorem~\ref{thmcurrentmink} are given by
\beq  \label{USymm20}
c_\beta := \partial^+_\omega  a(m_\beta^2) + \partial^+ _\omega b(m_\beta^2) + \partial^-_\omega  a(m_\beta^2) + \partial^-_\omega b(m_\beta^2)
\eeq
As explained above, even though the function $a+b$ is minimal at $m_\beta^2$,
it is in general not differentiable at this value. But the minimality implies that~$c_\beta \geq 0$.

The discontinuity of the derivatives of~$\hat{Q}$ on the mass shells implies that~$Q(x,y)$
will {\em{not}} decay rapidly as~$(y-x)^2 \rightarrow \pm \infty$. Instead, we obtain contributions which
decay only polynomially and oscillate on the {\em{Compton scale}}
(this oscillatory behavior comes about similar as explained for the Fourier transforms of the mass shells
in detail in~\cite[\S1.2.5]{cfs}).
Due to these oscillations on the Compton scale, the integrals in~\eqref{inttask}
are indeed well-defined, and the dominant contribution to the integrals will come from a
layer of width~$\sim m^{-1}$ around the hyperplane~$\{t=0\}$.
Therefore, although~$\L(x,y)$ does not decay rapidly,
the concept of the surface layer integral as introduced in Section~\ref{secsli} remains valid,
and the parameter~$\delta$ shown in Figure~\ref{fignoether1} can be identified
with the Compton scale~$\sim m_\alpha^{-1}$ of the Dirac particles. Thus the width of the
surface layer is a small but
macroscopic length scale. In particular, the surface layer integrals cannot be
identified with or considered as a generalization of the surface integrals of the classical Noether theorem.
However, in most situations of interest, when the surface is almost flat on the
Compton scale, the surface layer integral can be well-approximated by a corresponding surface integral.
Theorem~\ref{thmcurrentmink} shows that in the limiting case that the surface is a hyperplane,
the surface layer integral indeed goes over to a surface integral.

The just-mentioned oscillatory behavior of the integrand in~\eqref{inttask} implies that
the integrals will in general not exist in the Lebesgue sense. But they do exist in the sense
of an improper Riemann integral. For computational purposes, this is implemented
most conveniently by inserting convergence-generating factors.
We begin with the simplest possible choice of a convergence-generating factor~$e^{-\eta |t|}$.
Thus instead of~\eqref{inttask} we consider the integral
\beq
J = \lim_{\eta \searrow 0} \int_{t \geq 0} d^4x \int_{t < 0} d^4y \:e^{-\eta x^0 + \eta y^0}\: 
\im \big( \Sl \psi^u(y) \,|\, Q(y,x) \,\psi^u(x) \Sr \big) \label{Curreg1} \:.
\eeq

We now introduce a convenient representation for~$\hat{\psi}^u(k)$. Since the wave function~$\psi^u$ is
a linear combination of solutions of the Dirac equation corresponding to the masses~$m_\beta$
(with~$\beta=1,2,3$), its Fourier transform is supported on the mass shells~$k^2=m^2_\beta$.
Moreover, since in the Dirac sea vacuum all physical wave functions have negative frequency,
we can write~$\hat{\psi}^u(k) = (\hat{\psi}^u_\beta(k))_{\beta=1,2,3}$ as
\beq \label{USymm6}
\hat\psi^u_\beta(k) = 2 \pi\, \chi_\beta(\vec{k}) \: \delta \big( k^0 + \omega_{\beta,\vec k} \big)
\eeq
(with~$\omega_{\beta,\vec k}$ as in~\eqref{disperse}).
The Dirac equations~\eqref{Direqns} reduce to the algebraic equations
\beq \label{Diralg}
(\slashed{k}_\beta - m_\beta) \chi_\beta(\vec{k}) = 0 \qquad \text{where} \qquad
k_\beta := \big( -\omega_{\beta,\vec{k}}, \vec{k} \big)\:.
\eeq
The representation~\eqref{USymm6} has the convenient feature that the wave function at time~$t$
is given by
\[ \psi^u_\beta(t,\vec{x}) = \int \frac{d^4k}{(2 \pi)^4}\: \hat\psi^u_\beta(k)\: e^{- i k x}
= e^{i \omega_{\beta,\vec k} t} \int \frac{d^3k}{(2 \pi)^3}\: \chi_\beta(\vec k)\: e^{i \vec{k} \vec{x}} \:, \]
showing that~$\chi_\beta(\vec{k})$ simply is the {\em{spatial}} Fourier transform of
the Dirac wave function at time zero.

\begin{Lemma} \label{lemmaoffdiagonal}
The integral~\eqref{Curreg1}, can be written as
\begin{gather}
J = \sum_{\alpha, \beta=1}^3 J_{\alpha, \beta} \:, \label{Jsum} \\
\intertext{where the~$J_{\alpha, \beta}$ are given by}
J_{\alpha, \beta} = \lim_{\eta \searrow 0} \;\im \int \frac{d^4k}{(2 \pi)^4} \, \Sl  \chi_\alpha(\vec k ) \: \frac{ i}{k^0 + \omega_{\alpha,\vec k} + i \eta}  \:|\: \hat Q(k) \:  \chi_\beta(\vec k ) \: \frac{- i}{k^0 + \omega_{\beta,\vec k} - i \eta} \Sr \:. 
\label{JabDef}
\end{gather}
\end{Lemma}
\Proof We first rewrite~\eqref{Curreg1} as
\[ J = \lim_{\eta \searrow 0}\; \im \int d^4x \, \int d^4y \; \Sl \Theta(x^0)\,e^{-\eta x^0} \,  \psi^u(x) \,|\, Q(x,y)\, 
\Theta(-y^0) \, e^{\eta y^0} \,\psi^u(y) \Sr \:. \]
Since~$Q$ depends only on the difference vector~$y-x$, the~$y$-integration can be
regarded as a convolution in position space. We now rewrite this convolution as
a multiplication in momentum space. Setting
\[ \hat \psi_\eta^\pm (k) := \int \Theta_\eta (\pm y^0)\, \psi^u(y) \, e^{ i k y} \: d^4y \:, \]
where we introduced the ``regularized Heaviside function''
\[ \Theta_\eta(x) = \Theta(x)\, e^{-\eta x}\:, \]
we obtain
\[ J = \lim_{\eta \searrow 0}\;\im \int_{M} d^4x \,  \, \Sl \Theta_\eta(x^0) \,  \psi^u(x) \,|\, {\mathcal{F}}^{-1} \big( \hat{Q} \, \hat\psi_\eta^- \big)(x) \Sr \:, \]
where ${\mathcal{F}}^{-1}$ denotes the inverse Fourier transformation. Plancherel's theorem yields
\beq \label{USymm2}
J = \lim_{\eta \searrow 0}\;\im \int \frac{d^4k}{(2 \pi)^4} \, \Sl \hat \psi_\eta^+(k)  \,|\, \hat Q(k)
\,\hat \psi_\eta^-(k) \Sr \:.
\eeq

We next compute~$\hat \psi_\eta^\pm (k)$. Since multiplication in position space corresponds to
convolution in momentum space, we know that
\beq \label{hatppm}
\hat \psi_\eta^\pm (k) =  \int \frac{d\omega}{2 \pi} \:  \hat \Theta_\eta(\pm \omega) \: \hat \psi^u \big( k-(\omega,\vec 0) \big) \:.
\eeq
Here the Fourier transformation of the regularized Heaviside function is computed by
\beq \label{hatT}
\hat \Theta_\eta (\omega) = \int_{-\infty}^\infty \Theta_\eta(t) \: e^{i  \omega t}\: dt = \frac{i}{\omega + i \, \eta} \:.
\eeq

Using~\eqref{hatT} and~\eqref{USymm6} in~\eqref{hatppm}, we obtain
\[ \hat\psi_\eta^\pm ( k) = \bigg(\chi_\beta(\vec k ) \: \frac{i}{ \pm(k^0 + \omega_{\beta,\vec k}) + i \eta}
\bigg)_{\beta=1,2,3} \:. \]
Using these formulas in~\eqref{USymm2} gives the result.
\QED

The next lemma shows that the summands for~$\alpha \neq \beta$ drop out of~\eqref{Jsum}.
\begin{Lemma} The currents~\eqref{JabDef} satisfy the relation
\[ \sum_{\alpha \neq \beta} J_{\alpha, \beta} = 0 \:. \]
\end{Lemma}
\Proof
In the case~$\alpha \neq \beta$, we know that~$\omega_{\alpha,\vec k} \neq \omega_{\beta,\vec k}$,
so that in~\eqref{JabDef} there are two single poles at~$k^0 = -\omega_{\alpha, \vec{k}}-i\eta$
and~$k^0 = -\omega_{\beta, \vec{k}}+i\eta$. This makes it possible to take the limit~$\eta \rightarrow 0$
using the formula
\[ \lim_{\eta \searrow 0} \frac{1}{x \pm i \eta} = \frac{\text{PP}}{x} \mp i \pi \, \delta(x) \]
(where PP denotes the principal value). We thus obtain
\begin{align*}
J_{\alpha, \beta} \;=& -\im \int_{M} \frac{d^4k}{(2 \pi)^4} \, 
\frac{\text{PP}}{k^0 + \omega_{\alpha,\vec k}}  \:\frac{\text{PP}}{k^0 + \omega_{\beta,\vec k}} \;
\Sl  \chi_\alpha(\vec k ) \:|\: \hat Q(k) \:  \chi_\beta(\vec k ) \Sr \\
&- \im \int_{M} \frac{d^4k}{(2 \pi)^4} \, \Sl  \chi_\alpha(\vec k ) \: \big(- i \pi\, \delta(k^0 + \omega_{\alpha,\vec k}) \big) \:|\:  \hat Q(k) \: \chi_\beta(\vec k ) \: \frac{\text{PP}}{k^0 + \omega_{\beta,\vec k}} \Sr \\
&- \im \int_{M} \frac{d^4k}{(2 \pi)^4} \, \Sl  \chi_\alpha(\vec k ) \: \frac{\text{PP}}{k^0 + \omega_{\alpha,\vec k}}  \:|\: \hat Q(k) \:  \chi_\beta(\vec k ) \: \big(i \pi\, \delta(k^0 + \omega_{\beta,\vec k}) \big) \Sr \:.
\end{align*}
Carrying out the $k^0$-integration in the last two lines gives
\begin{align}
J_{\alpha, \beta} \;=& -\im \int_{M} \frac{d^4k}{(2 \pi)^4} \, 
\frac{\text{PP}}{k^0 + \omega_{\alpha,\vec k}}  \:\frac{\text{PP}}{k^0 + \omega_{\beta,\vec k}} \;
\Sl  \chi_\alpha(\vec k ) \:|\: \hat Q(k) \:  \chi_\beta(\vec k ) \Sr \notag \\
&+\pi \re \int_{M} \frac{d^3k}{(2 \pi)^4} \, \Sl  \chi_\alpha(\vec k ) \:  |\:  \hat
Q \big( -\omega_{\alpha,\vec k}, \vec{k} \big) \: \chi_\beta(\vec k ) \: \frac{\text{PP}}{-\omega_{\alpha,\vec k} + \omega_{\beta,\vec k}} \Sr \notag \\
&+\pi \re \int_{M} \frac{d^3k}{(2 \pi)^4} \, \Sl  \chi_\alpha(\vec k ) \: \frac{\text{PP}}{-\omega_{\beta,\vec k} + \omega_{\alpha,\vec k}}  \:|\: \hat Q\big( -\omega_{\beta,\vec k}, \vec{k} \big) \:  \chi_\beta(\vec k ) \Sr \notag \\
=& -\im \int_{M} \frac{d^4k}{(2 \pi)^4} \, 
\frac{\text{PP}}{k^0 + \omega_{\alpha,\vec k}}  \:\frac{\text{PP}}{k^0 + \omega_{\beta,\vec k}} \;
\Sl  \chi_\alpha(\vec k ) \:|\: \hat Q(k) \:  \chi_\beta(\vec k ) \Sr \label{aneqb1} \\
&+\pi \re \int_{M} \frac{d^3k}{(2 \pi)^4} \:\frac{\text{PP}}{\omega_{\alpha,\vec k} - \omega_{\beta,\vec k}} \notag \\
&\qquad \qquad \quad \times \Sl  \chi_\alpha(\vec k ) \:  |\:  \Big( \hat{Q} \big( -\omega_{\beta,\vec k}, \vec{k} \big) - \hat{Q} \big( -\omega_{\alpha,\vec k}, \vec{k} \big) \Big) \chi_\beta(\vec k ) \: \Sr \:. \label{aneqb2}
\end{align}
Obviously, the contribution~\eqref{aneqb1} is anti-symmetric when exchanging~$\alpha$ and~$\beta$.
In the contribution~\eqref{aneqb2}, on the other hand,
we can use the Dirac equation~\eqref{Diralg} together with~\eqref{62f} to rewrite
the spin scalar product as
\[ \Sl  \chi_\alpha(\vec k ) \:  |\:  \big( (a+b)(m_\beta^2) - (a+b)(m_\alpha^2) \big) \chi_\beta(\vec k ) \: \Sr \:, \]
and this vanishes by~\eqref{abmin}. This gives the result.
\QED

Using this lemma, our conserved integral~\eqref{Jsum} simplifies to
\beq \label{Jsum2}
J = \sum_{\beta=1}^3 J_{\beta, \beta} \:.
\eeq
We now compute~$J_{\beta, \beta}$. First,
\begin{align}
& J_{\beta, \beta}
= \lim_{\eta \searrow 0} \;\im \int_{M} \frac{d^4k}{(2 \pi)^4} \, \Sl  \chi_\beta(\vec k ) \: \frac{ i}{k^0 + \omega_{\beta,\vec k} + i \eta}  \:|\: \hat Q(k) \:  \chi_\beta(\vec k ) \: \frac{- i}{k^0 + \omega_{\beta,\vec k} - i \eta}   \Sr \nonumber \\
&\:= \lim_{\eta \searrow 0} \int_{M} \frac{d^4k}{(2 \pi)^4} \, \Sl  \chi_\beta(\vec k ) \:|\: \hat Q(k) \: \frac{1}{2i} \bigg( \frac{-1}{( k^0 + \omega_{\beta,\vec k} - i \eta)^2} - \frac{-1}{( k^0 + \omega_{\beta,\vec k} + i \eta)^2} \bigg)
\chi_\beta(\vec k ) \: \Sr \nonumber  \\
&\,\overset{(\star)}{=} -\lim_{\eta \searrow 0} \int \frac{d^3k}{(2 \pi)^2} \int_{-\infty}^\infty \frac{dq}{2 \pi} \, \Sl 
\chi_\beta(\vec k ) \: |\: \hat Q 
\big(q - \omega_{\beta,\vec k}, \vec k \big) \: \frac{1}{2i} \bigg( \frac{1}{( q + i \eta)^2} - \frac{1}{( q - i \eta)^2} \bigg)
\chi_\beta(\vec k ) \: \Sr \nonumber  \\
&\:= - 2 \lim_{\eta \searrow 0} \int \frac{d^3k}{(2 \pi)^3} \, \Sl  \chi_\beta(\vec k ) \: |\: \int_{-\infty}^\infty \frac{dq}{2 \pi} \:\bigg( \hat Q\big(q - \omega_{\beta,\vec k}, \vec k \big) \;  \frac{q \, \eta}{( q^2 +\eta^2)^2} \bigg)  \chi_\beta(\vec k ) \: \Sr \:,  \label{CurrentCons1}
\end{align}
where in~$(\star)$ we introduced the variable~$q=k^0 + \omega_{\beta,\vec k}$.
We now use~\eqref{semidiff} to expand~$\hat{Q}$ for small~$q$ according to
\begin{align*}
&\hat{Q}\big( q- \omega_{\beta,\vec k} , \vec k\big) \\
&=  \, \hat Q\big(- \omega_{\beta,\vec k} , \vec k \big)
+ q  \, \Theta(q) \:\partial^+_\omega \hat Q\big( - \omega_{\beta,\vec k} , \vec k \big)
+ q \, \Theta(-q)  \: \partial^-_\omega\hat Q\big( - \omega_{\beta,\vec k} , \vec k \big) + o(q)
\end{align*}
(where $o(q)$ is the usual remainder term).
Substituting this Taylor expansion into~\eqref{CurrentCons1}, the constant term
of the expansion drops out because the integrand is odd.
For the left and right derivatives, the integral can be carried out explicitly using that
\beq \label{exint}
\int_0^\infty \frac{q^2 \,\eta}{(q^2+\eta^2)^2} \, dq = \frac{\pi}{4}
= \int_{- \infty}^0 \frac{q^2 \,\eta}{(q^2+\eta^2)^2} \, dq \:.
\eeq
Thus, disregarding the remainder term, we obtain
\beq \label{USymm10}
J_{\beta, \beta}= - \frac{1}{4} \int \frac{d^3k}{(2 \pi)^3} \, \Sl  \chi_\beta(\vec k ) \: |\: 
\Big( (\partial^+_\omega + \partial^-_\omega) \hat Q\big( - \omega_{\beta,\vec k} , \vec k \,\big) \Big)
 \chi_\beta(\vec k ) \: \Sr \:.
\eeq
This formula corresponds to the result of Lemma~\ref{lemmagen} 
in our setting where~$\hat{Q}(k)$ is not differentiable on the mass shells.

It remains to analyze the remainder term. Naively, the integrated remainder term is of the order~$\eta$ and
should thus vanish in the limit~$\eta \searrow 0$. This could indeed be proved if
we knew for example that the
function~$\hat{Q}( \,.\, - \omega_{\beta,\vec k}, \vec k )$ is integrable.
However, since~$\hat{Q}$ is only defined on the lower mass cone (see Definition~\ref{def611}),
such arguments cannot be applied. Our method for avoiding this technical problem is to
work with a convergence-generating factor with compact support in momentum space. To this end,
we choose a non-negative test function~$\hat{g} \in C^\infty_0((-1,1))$ with~$\hat{g}(-\omega)=\hat{g}(\omega)$
for all~$\omega \in \R$ and~$\int_\R \hat{g}(\omega)\, d\omega =2 \pi$.
For given~$\sigma >0$ we set
\[ \hat{g}_\sigma(\omega) = \frac{1}{\sigma}\: \hat{g} \Big( \frac{\omega}{\sigma} \Big) \qquad \text{and} \qquad
g_\sigma(t) = \int_{-\infty}^\infty \frac{d\omega}{2 \pi}\: \hat{g}_\sigma(\omega)\: e^{-i \omega t} \:. \]
In the limit~$\sigma \searrow 0$, the functions~$g_\sigma(t)$ go over to the constant function one.

\begin{Lemma} \label{lemmacur1} Replacing~\eqref{Curreg1} by
\beq \label{Jnew}
J = \lim_{\sigma \searrow 0} \int_{t \geq 0} d^4x \int_{t < 0} d^4y \:g_\sigma(x^0) \: g_\sigma(y^0)\:
\im \big( \Sl \psi^u(y) \,|\, Q(y,x) \,\psi^u(x) \Sr \big) \:,
\eeq
the resulting function~$J$ is of the form~\eqref{Jsum2} with~$J_{\beta, \beta}$
as given by~\eqref{USymm10}.
\end{Lemma}
\Proof Again rewriting~\eqref{Jnew} in momentum space and using that~$\hat{g}$ has compact support,
one sees that the resulting integrand of~$J_{\alpha, \beta}$ is well-defined for any~$\vec{k}$
for sufficiently small~$\sigma$.
In order to relate the functions~$g_\sigma$ in~\eqref{Jnew} to the
factor~$e^{-\eta x^0 + \eta y^0}$ in~\eqref{Curreg1}, it is most convenient to work with
the Laplace transform. Thus we represent the functions~$g_\sigma$ in~\eqref{Jnew}
for~$x^0>0$ and~$y^0<0$ as
\[ g_\sigma(x^0) = \frac{1}{\sigma} \int_0^\infty h\Big( \frac{\eta}{\sigma} \Big)\, e^{-\eta x^0}\: d\eta
\qquad \text{and} \qquad
g_\sigma(y^0) = \frac{1}{\sigma} \int_0^\infty h\Big( \frac{\tilde{\eta}}{\sigma} \Big)\, e^{\tilde{\eta} y^0}\:
d\tilde{\eta} \:, \]
where~$h$ is the inverse Laplace transform of~$g$
(for basics on the Laplace transform see for example~\cite{davies}).
A straightforward computation shows that the result of Lemma~\ref{lemmaoffdiagonal}
remains valid with the obvious replacements.
The computation of~$J_{\beta, \beta}$, on the other hand, needs to be modified as follows. 
Formula~\eqref{CurrentCons1} remains valid after the replacement
\[ \lim_{\eta \searrow 0} \;\cdots\; \frac{q \, \eta}{( q^2 +\eta^2)^2} \;\longrightarrow\;
 \lim_{\sigma \searrow 0} \frac{1}{\sigma^2} \int_0^\infty h\Big( \frac{\eta}{\sigma} \Big)\:d\eta
\int_0^\infty h\Big( \frac{\tilde{\eta}}{\sigma} \Big)\: d\tilde{\eta}
\;\cdots\; \frac{q \, (\eta+\tilde{\eta})}{2 ( q^2 +\eta^2) ( q^2 +\tilde{\eta}^2)} \:. \]
Substituting the Taylor expansion of~$\hat{Q}$, the first integral in~\eqref{exint}
is to replaced by the integral
\[ \int_0^\infty \frac{q^2 \, (\eta+\tilde{\eta})}{2 ( q^2 +\eta^2) ( q^2 +\tilde{\eta}^2)}  = \frac{\pi}{4} \]
(and similarly for the second integral in~\eqref{exint}).
In this way, one again obtains~\eqref{USymm10}, but now the remainder term vanishes in
the limit~$\sigma \searrow 0$. 
\QED

We now compute~$J_{\beta, \beta}$ more explicitly.
\begin{Lemma} \label{lemmacur2}
The currents~$J_{\beta, \beta}$ given by~\eqref{USymm10} can be written as
\beq \label{Jfinal}
J_{\beta, \beta} = - \frac{m_\beta \,c_\beta}{2} \int_{\R^3} \Sl \psi_\beta^u(x) | \gamma^0 \psi_\beta^u(x) \Sr\:
d^3x
\eeq
with the constants~$c_\beta$ as in~\eqref{USymm20}.
\end{Lemma}
\Proof Using~\eqref{62f} and applying the chain rule for semi-derivatives, we obtain
\beq \label{USymm3}
\partial^\pm_\omega \, \hat Q \Big( - \omega_{\beta,\vec k}, \vec k \Big) 
= - 2 \omega_{\beta,\vec k} \Big(\partial^\pm_\omega  a(k_-^2) \frac{\slashed{k}_-}{| k_- | } + \partial^\pm_\omega b(k_-^2) \Big) + a(k_-^2) \, \frac{\partial}{\partial k^0} \Big( \frac{\slashed{k} }{|k|} \Big) \Big|_{k=k_-},
\eeq
where we set~$k_- = (- \omega_{\beta,\vec k} , \vec k )$ and~$|k_-| = \sqrt{k_-^2}=m_\beta$.
This formula can be further simplified when taking the expectation value with the spinor~$\chi_\beta(\vec{k})$:
In the last summand in~\eqref{USymm3}, we first compute the~$k$-derivative,
\[ \frac{\partial}{\partial k^0} \frac{\slashed{k} }{|k|} \Big|_{k = k_-} = \frac{\gamma^0}{m_\beta} 
- \frac{\slashed{k}_-}{|k_-|^3} \, k_-^0 \:. \]
Taking the expectation value with the spinor~$\chi_\beta(\vec{k})$ and using the Dirac equation
\[ (\slashed{k}_- - m_\beta) \chi_\beta(\vec{k})=0\:, \]
we obtain the relations
\begin{gather}
\Sl \chi_\beta(\vec{k}) | \slashed{k}_-\, \chi_\beta(\vec{k}) \Sr = m_\beta\: \Sl \chi_\beta(\vec{k}) | \chi_\beta(\vec{k}) \Sr \\
2 m_\beta \;\Sl \chi_\beta(\vec{k}) | \gamma^0 \chi_\beta(\vec{k}) \Sr
= \Sl \chi_\beta(\vec{k}) | \big\{ \slashed{k}_-, \gamma^0 \big\} \chi_\beta(\vec{k}) \Sr
= -2 \omega_{\beta,\vec k} \, \Sl \chi_\beta(\vec{k}) | \chi_\beta(\vec{k}) \Sr \:. \label{rel2}
\end{gather}
In this way, the last summand in~\eqref{USymm3} gives zero.
In the remaining first summand in~\eqref{USymm3}, we again employ the Dirac
equation~$(\slashed{k}_- - m_\beta) \chi_\beta(\vec{k})=0$ to obtain
\begin{align*}
\bigg(\partial^\pm_\omega  a(k_-^2) \frac{\slashed{k}_-}{| k_- | }
+ \partial^\pm_\omega b(k_-^2) \bigg) \chi_\beta(\vec{k}) =
\partial^\pm_\omega  \Big(a(k_-^2) + b(k_-^2) \Big) \chi_\beta(\vec{k}) \:.
\end{align*}
We conclude that
\beq  \label{USymm19}
J_{\beta, \beta} = \frac{1}{2} \, c_\beta    \int \frac{d^3k}{(2 \pi)^3} \,\omega_{\beta,\vec k}  \:
\Sl \chi_\beta(\vec k) |  \chi_\beta(\vec k ) \Sr
\eeq
with~$c_\beta$ as in~\eqref{USymm20}.
We finally use~\eqref{rel2} and apply Plancherel's theorem.
\QED

Combining Lemma~\ref{lemmacur1} and Lemma~\ref{lemmacur2} 
gives the conservation law~\eqref{thmcurr1}. This concludes
the proof of Theorem~\ref{thmcurrentmink}.

\subsection{Clarifying Remarks} \label{secremark}
The following remarks explain and clarify various aspects of
the above constructions and results.
\begin{Remark} \label{remdiscuss} {\bf{(differentiability of variations)}} {\em{ 
We now explain in which sense the the differentiability assumption 
on the function~$\ell \circ \Phi$ in Theorem~\ref{thmcurrent} is satisfied.
First, the above computations show that, working with the specific form of~$\hat{Q}$ in the continuum limit,
the $\tau$-derivative exists and is finite.
However, this does not necessarily imply that for any UV regularization,
the corresponding local minimizers~$(\H, \F, \rho^\varepsilon)$
also satisfy the differentiability assumptions on the function~$\ell \circ \Phi$ in
Theorem~\ref{thmcurrent}. Indeed, thinking of a lattice regularization,
we expect that the function~$\ell \circ \Phi$ with~$\Phi$ according to~\eqref{varunit}
and~\eqref{Utaudef} will typically {\em{not}} be continuously differentiable in~$\tau$
(because in this case, $\ell$ is a sum of terms involving the Lagrangian, which is only
Lipschitz continuous). In order to bypass this technical problem, 
for a given local minimizer~$(\H, \F, \rho^\varepsilon)$
one can modify~$\Phi$ such as to obtain a variation~$\Phi^\varepsilon$ for which the
function~$\ell \circ \Phi^\varepsilon$
is continuously differentiable in~$\tau$ (for details see~\cite{jet}).
For this modified variation, we have the conservation law of Theorem~\ref{thmcurrent}.
The strategy is to choose the~$\Phi^\varepsilon$ for every~$\varepsilon>0$
in such a way that in the limit~$\varepsilon \searrow 0$, the variations converge
in a suitable weak topology to the variation~$\Phi_\tau$ as given by~\eqref{varunit}
and~\eqref{Utaudef}. In non-technical terms, we modify~$\Phi_\tau$ by ``microscopic fluctuations''
in such a way that the functions~$\ell \circ \Phi^\varepsilon$ become differentiable in~$\tau$
for all~$\varepsilon>0$. In the limit~$\varepsilon \searrow 0$, the microscopic fluctuations
should drop out to give Theorem~\ref{thmcurrentmink}.

At present, this procedure cannot be carried out because, so far,
no local minimizers~$(\H, \F, \rho^\varepsilon)$ have been constructed which describe
regularized Dirac sea configurations. The difficulty is to arrange the regularization in such a way
that the EL equations are satisfied without error terms. A first step towards the construction of such ``optimal
regularizations'' is given in~\cite{reg}. 
\QEDrem }}
\end{Remark}

\begin{Remark} {\bf{(weight factors)}} \label{remweights} {\em{
As explained in~\cite[Section~2 and Appendix~A]{reg}, one may introduce positive
weight factors~$\rho_\beta$ into the ansatz~\eqref{Pvac},
\[ P(x,y) = \sum_{\beta=1}^3 \rho_\beta \int \frac{d^4k}{(2 \pi)^4}\: (\slashed{k}+m_\beta)\:
\delta \big(k^2-m_\beta^2 \big)\: e^{-ik(x-y)}\:. \]
The above analysis immediately extends to this situation simply by inserting suitable factors of~$\rho_\beta$
into all equations. In particular, the resulting conserved quantity~\eqref{Jfinal} becomes
\[ J_{\beta, \beta} = - \frac{\rho_\beta\, m_\beta \,c_\beta}{2} \int_{\R^3} \Sl \psi_\beta^u(x) | \gamma^0
\psi_\beta^u(x) \Sr\: d^3x \:. \]
Consequently, the conserved current in~\eqref{thmcurr1} is to be modified to
\[ \sum_{\beta=1}^3 \rho_\beta\, m_\beta\, c_\beta \int_{t=\text{const}} \!\!\!\!\!\!\!\!\Sl \psi_\beta^u(x)
| \gamma^0 \psi_\beta^u(x) \Sr\:d^3x\:. \]
The role of the weight factors in the interacting case will be explained in the next remark.
\QEDrem }} \end{Remark}

\begin{Remark} {\bf{(interacting systems)}} \label{reminteract} {\em{
We point out that for the derivation of Theorem~\ref{thmcurrentmink}, we
worked with the vacuum Dirac equations~\eqref{Direqns}, so that no
interaction is present. In particular, the generations have an independent dynamics,
implying that current conservation holds separately for each generation, i.e.
\beq \int_{t=t_0} \!\!\!\!\Sl \psi_\beta^u(x) | \gamma^0 \psi_\beta^u(x) \Sr\:d^3x
= \int_{t=t_1} \!\!\!\!\Sl \psi_\beta^u(x) | \gamma^0 \psi_\beta^u(x) \Sr\:d^3x \quad
\text{for all~$\beta=1,2,3$}\:. \label{conssep}
\eeq
Let us now discuss the typical situation of a scattering process in which the
Dirac equations~\eqref{Direqns} only hold asymptotically as~$t \rightarrow \pm \infty$.
In this case, choosing~$\Omega$ so large that it contains the interaction region,
one can compute the surface layer integrals again for the free Dirac equation
to obtain the conservation law~\eqref{thmcurr1}, where~$t_0$ lies in the past
and~$t_1$ in the future of the interaction region.
In this way, the conservation law of Theorem~\ref{thmcurrentmink}
immediately extends to interacting systems.

In this interacting situation, current conservation no longer holds for each
generation separately (thus~\eqref{conssep} is violated). Instead,
as a consequence of the Dirac dynamics, only the total charge
\beq \label{totcharge}
\sum_{\beta=1}^3
\int_{t=\text{const}} \!\!\!\!\!\!\!\!\Sl \psi_\beta^u(x) | \gamma^0 \psi_\beta^u(x) \Sr\:d^3x
\eeq
is conserved.
In order for this conservation law to be compatible with~\eqref{thmcurr1}, we need
to impose that
\beq \label{cc1}
m_\alpha\, c_\alpha = m_\beta\, c_\beta \qquad \text{for all~$\alpha,\beta=1,2,3$}\:.
\eeq
This is a mathematical consistency condition which gives information on the possible
form of the distribution~$\hat{Q}(k)$ in the continuum limit (as specified in
Definition~\ref{def611} above).
If weight factors are present (see Remark~\ref{remweights} above), this consistency condition
must be modified to
\beq \label{cc2}
\rho_\alpha\, m_\alpha\, c_\alpha = \rho_\beta\, m_\beta\, c_\beta \qquad \text{for all~$\alpha,\beta=1,2,3$}\:.
\eeq
The conditions~\eqref{cc1} and~\eqref{cc2} are crucial for the future project of
extending the state stability analysis in~\cite{vacstab} to systems involving neutrinos.
\QEDrem }} \end{Remark}

\begin{Remark} {\bf{(normalization of the fermionic projector)}} \label{remnorm} {\em{
The conservation law of Theorem~\ref{thmcurrentmink} has an important implication
for the normalization of the fermionic projector, as we now explain.
As worked out in detail in~\cite{norm}, there are two alternative normalization methods 
for the fermionic projector: the spatial normalization and the mass normalization.
In~\cite[Section~2.2]{norm} the advantages of the spatial normalization are discussed,
but no decisive argument in favor of one of the normalization methods is given.
Theorem~\ref{thmcurrentmink} decides the normalization problem in favor of the
spatial normalization. Namely, this theorem shows that the dynamics as described by the causal
action principle gives rise to a conservation law which in the continuum limit reduces
to the spatial integrals~\eqref{thmcurr1}. As explained in Remark~\ref{remweights} above,
the mathematical consistency to the Dirac dynamics implies that~\eqref{thmcurr1}
coincides with the conserved total charge~\eqref{totcharge}.
The resulting conservation law is compatible with the spatial normalization, but contradicts
the mass normalization. We conclude that the spatial normalization of the fermionic projector
is indeed the correct normalization method which reflects the intrinsic conservation laws of the
causal fermion system.
\QEDrem }} \end{Remark}

\section{Example: Conservation of Energy-Momentum} \label{secexEM}
The conservation laws in Theorem~\ref{thmsymmgis2} also give rise to the
{\em{conservation of energy and momentum}}, as will be worked out in this section.

\subsection{Generalized Killing Symmetries and Conservation Laws}
In the classical Noether theorem, the conservation
laws of energy and momentum
are a consequence of space-time symmetries described most conveniently with
the notion of Killing fields. Therefore, one of our tasks is to extend this notion to the setting of causal
fermion systems. In preparation, we recall the procedure in the classical Noether theorem
from a specific point of view:
In the notion of a Killing field, one distinguishes the background geometry from the
additional particles and fields. The background geometry must have a symmetry
as described by the Killing equation. The additional particles and fields, however, do not
need to have any symmetries. Nevertheless, one can construct a symmetry of the whole system
by actively transporting the particles and fields along the flow lines of the Killing field.
The conservation law corresponding to this symmetry transformation gives rise to
the conservation of energy and momentum.

In a causal fermion system, there is no clear-cut distinction between the background geometry
and the particles and fields of the system, because all of these structures are inherent in the
underlying causal fermion system and mutually influence each other via the causal action principle.
Therefore, instead of working with a symmetry of the background geometry,
we shall work with the notion of an approximate symmetry. By actively transforming
those physical wave functions which do not respect the symmetry,
such an approximate symmetry again gives rise to an exact
symmetry transformation, to which our Noether-like theorems apply.

More precisely, one begins with a $C^1$-family of transformations~$(f_\tau)_{\tau \in(-\tau_{\max}, \tau_{\max})}$ of space-time,
\[ f_\tau \::\: M \rightarrow M \qquad \text{with} \qquad f_0 = \1 \:, \]
which preserve the universal measure in the sense that~$(f_\tau)_* \rho = \rho$.
This family can be regarded as the analog of the flow in space-time along a classical Killing field.
Moreover, one considers a family of unitary transformations~$(\scrU_\tau)_{\tau \in (-\tau_{\max}, \tau_{\max}) }$
on~$\H$ with the property that
\beq \label{PropU}
\scrU_{-\tau} \,\scrU_\tau = \1 \qquad \text{for all~$\tau \in (-\tau_{\max}, \tau_{\max})$}\:, \eeq
and defines the variation
\beq \label{EMEq3}
\Phi \::\: (-\tau_{\max}, \tau_{\max}) \times M \rightarrow \F \:,\qquad
\Phi(\tau,x) := \scrU_\tau \, x \, \scrU^{-1}_\tau \:.
\eeq
Combining these transformations should give rise to an
{\em{approximate symmetry}} of the wave evaluation operator~\eqref{weo}
in the sense that if we compare the transformation of the space-time point with the unitary transformation
by setting
\beq \label{Edef}
E_\tau(u,x) := (\Psi u)\big(f_\tau(x) \big) - \big( \Psi \scrU^{-1}_\tau u \big)(x) \qquad (x \in M, u \in \H) \:,
\eeq
then the operator~$E_\tau : \H \rightarrow C^0(M, SM)$ should be so small
that the first variation is well-defined in the continuum limit (for details see Section~\ref{CorrDiracEM} below).
There are various ways in which this smallness condition could be
formulated. We choose a simple method which is most convenient for our purposes.

\begin{Def} \label{defkilling}
The transformation $(f_\tau)_{\tau \in (-\tau_{\max},\tau_{\max})}$ is called a {\bf{Killing symmetry with finite-dimensional support}} of the causal fermion system if it is a symmetry of the universal measure that preserves the trace (see Definition~\ref{defsymmnc}) and if there exists a finite-dimensional subspace $K \subset \H$ and 
a family of unitary operators~$(\scrU_\tau)_{\tau \in (-\tau_{\max},\tau_{\max})}$ with the property~\eqref{PropU} such that
\beq \label{CondSmall}
E_\tau(u,x) = 0 \qquad \text{for all~$u \in K^\perp$ and~$x \in M$}\:.
\eeq
\end{Def}

We now formulate a general conservation law.
\begin{Thm} \label{ThmKillingCons} Let $\rho$ be a local minimizer (see Definition~\ref{deflocmin}) and $(f_\tau)_{\tau \in (-\tau_{\max},\tau_{\max})}$ be a Killing symmetry of the causal fermion system. Then the following conservation law holds:
\begin{align} \label{KillingConserve} \begin{split}
\frac{d}{d\tau} \int_\Omega d\rho(x) \int_{M \setminus \Omega} d\rho(y) \:\Big( &\L_\kappa \big( f_\tau(x), y\big) -  \L_\kappa \big(x, f_\tau( y)\big) \\
&-\L_\kappa \big( \Phi_\tau(x), y\big) + \L_\kappa \big( x,\Phi_\tau( y) \big) \Big) \Big|_{\tau=0} = 0 \:.
\end{split}
\end{align}
\end{Thm}
\Proof Again using Lemma~\ref{PhiSymmLag}, we know that the
variation~\eqref{EMEq3} is a symmetry of the Lagrangian. Hence
\begin{align*}
\int_M d\rho(x) \int_\Omega d\rho(y) \, \L_\kappa \big(\Phi_\tau(x),y \big) =&  \int_M d\rho(x) \int_\Omega d\rho(y) \, \L_\kappa \big( x, \Phi_{-\tau}(y) \big) \\
=&  \int_\Omega d\rho(y) \: \ell \big(\Phi_\tau (y) \big) \:.
\end{align*}
Using this equation in Proposition~\ref{prpuseful2}, we obtain
\beq \label{phirel}
0 = \frac{d}{d \tau}  \int_\Omega d\rho(x) \, \int_{M\setminus \Omega} d\rho(y)\; \Big( \L_\kappa \big(\Phi_\tau(x),y
\big) - \L \big( x,\Phi_\tau(y) \big) \Big) \Big|_{\tau = 0} \:.
\eeq

For the transformations~$f_\tau$, on the other hand, we have the relations
\[ \int_M \L_\kappa \big(f_\tau(x) , y \big) \: d\rho(x) = \int_\F \L_\kappa(z,y)\: d\big( (f_\tau)_\ast \rho \big) (z) 
= \int_M  \L_\kappa(x,y) \:d\rho(x) \:, \]
where in the last step we used that $f_\tau$ is a symmetry of the universal measure. Since $f_\tau$ also
preserves the trace, it is a generalized integrated symmetry (see Definition~\ref{defgis2}). Applying
Theorem~\ref{thmsymmgis2}, we obtain
\beq \label{frel}
\frac{d}{d\tau} \int_\Omega d\rho(x) \int_{M \setminus \Omega} d\rho(y) \:\Big( \L_\kappa\big( f_\tau(x), y\big) -
\L_\kappa \big( x, f_\tau(y) \big) \Big) \Big|_{\tau=0} = 0 \:.
\eeq

Subtracting~\eqref{frel} from~\eqref{phirel} gives the result.
\QED

We remark that the vector field~$w := \delta f$ is tangential to~$M$ and describes
a transformation of the space-time points. The variation~$\delta \Phi$, on the other hand,
is a vector field in~$\F$ along~$M$. It will in general not be tangential to~$M$.
The difference vector field~$v := w - \delta \Phi$ 
can be understood as an active transformation of all the objects in space-time which
do not have the space-time symmetry (similar to the parallel transport of the particles and fields along the flow lines of the Killing field in the classical Noether theorem as described above).
The variation of the integrand in~\eqref{KillingConserve} can be rewritten as a variation
in the direction~$v$; for example,
\[ \frac{d}{d\tau} \Big( \L_\kappa \big( f_\tau(x), y\big) -\L_\kappa \big( \Phi_\tau(x), y\big) \Big) \Big|_{\tau=0}
= \delta_{v(x)} \L_\kappa(x,y) \:. \]
Expressing~$v$ in terms of the operator~$E$ in~\eqref{Edef} and using~\eqref{CondSmall}
will show that~$v$ is indeed so small (in a suitable sense) that the 
corresponding variation of the Lagrangian will be well-defined and finite.

\subsection{Correspondence to the Dirac Energy-Momentum Tensor} \label{CorrDiracEM}
In order to get the connection to the conservation of energy and momentum,
as in Section~\ref{seccurcor} we consider the vacuum Dirac equation and the limiting case
that~$\Omega$ exhausts the region between two Cauchy surfaces~$t=t_0$ and~$t=t_1$
(see Figure~\ref{fignoether2}). Recall that the energy-momentum tensor of a Dirac wave function~$\psi$
is given by
\[ T_{jk} = \frac{1}{2}\:
\re  \left( \Sl  \psi |   \, \gamma_j\, i \partial_k \psi \Sr + \Sl  \psi |   \, \gamma_k\, i \partial_j \psi \Sr \right)
= - \im  \Sl  \psi |   \, \gamma_{(j}\, \partial_{k)} \psi \Sr \:. \]
We consider the situation of the vacuum Dirac sea with a finite number of holes
describing the anti-particle states~$\phi_1, \ldots, \phi_{\na}$
(for the description of particle states see again Remark~\ref{remmicro}).
The effective energy-momentum tensor is minus the sum of the energy-momentum tensors
of all the anti-particle states. Thus for a fixed value of the generation index~$\beta$, we set
\[ (T_\beta)_{jk} = \sum_{i=1}^{\na} \im  \Sl  \phi_{i,\beta} |   \, \gamma_{(j}\, \partial_{k)} \phi_{i, \beta} \Sr \:. \]
In order to treat the generations, as in Theorem~\ref{thmcurrentmink} we take a linear combination
involving the non-negative constants~$c_\beta$ introduced in~\eqref{USymm20}.

\begin{Thm} {\bf{(energy conservation)}} \label{thmEMcons}
Let~$(\H, \F, \rho^\varepsilon)$ be local minimizers of the causal action
describing the Minkowski vacuum~\eqref{Pvac} together with particles and anti-particles.
Considering the limiting procedure explained in Figure~\ref{fignoether2} and taking the
continuum limit, the conservation law of Theorem~\ref{ThmKillingCons} goes over to
\[ \sum_{\beta=1}^3 m_\beta c_\beta \int_{t=t_0} d^3x \, (T_\beta)^0_0\: d^3x = 
\sum_{\beta=1}^3 m_\beta c_\beta \int_{t=t_1} d^3x \, (T_\beta)^0_0\: d^3x \:. \]
\end{Thm}

This theorem can be extended immediately to energy-momentum conservation
on general Cauchy surfaces:
\begin{Corollary} {\bf{(energy-momentum conservation on Cauchy surfaces)}} \label{corEMcons} \\
Let~$\scrN_0, \scrN_1$ be two Cauchy surfaces in Minkowski space, where~$\scrN_1$
lies to the future of~$\scrN_0$.
Then, under the assumptions of Theorem~\ref{thmEMcons},
the conservation law of Theorem~\ref{ThmKillingCons} goes over to the conservation law for
the energy and momentum integrals
\beq \label{Tfinal}
\sum_{\beta=1}^3 m_\beta\, c_\beta \int_{\scrN_0} (T_\beta)^j_k \,\nu_j\:d\mu_{\scrN_0}
= \sum_{\beta=1}^3 m_\beta\, c_\beta \int_{\scrN_1} (T_\beta)^j_k \,\nu_j \:d\mu_{\scrN_1} \:,
\eeq
where~$\nu$ again denotes the future-directed normal and~$k \in \{0,\ldots, 3\}$.
\end{Corollary}
\Proof We use similar arguments as in the proof of Corollary~\ref{corcurrent}.
More precisely, the conservation of classical energy implies that
\[ \int_{t=t_0} d^3x \, (T_\beta)^0_0\: d^3x = \int_{\scrN} (T_\beta)^j_0 \,\nu_j\: d\mu_{\scrN} \:. \]
This gives~\eqref{Tfinal} in the case~$k=0$.
Applying a Lorentz boost, we obtain
\[ \sum_{\beta=1}^3 m_\beta\, c_\beta \int_{\scrN_0} (T_\beta)^j_k \,\nu_j\: K^k\:d\mu_{\scrN_0}
= \sum_{\beta=1}^3 m_\beta\, c_\beta \int_{\scrN_1} (T_\beta)^j_k \,\nu_j \:K^k\:d\mu_{\scrN_1} \:, \]
where~$K$ is the Killing field obtained by applying the Lorentz boost to the vector field~$\partial_t$.
This gives the result.
\QED
These results shows that the conservation laws of energy and momentum correspond to more general
conservation laws in the setting of causal fermion systems.

The remainder of this section is devoted to the proof of Theorem~\ref{thmEMcons}.
Let~$(\H, \F, \rho^\varepsilon)$ be a regularized vacuum Dirac sea configuration
together with anti-particles (for details see~\cite[Sections~1.2 and~3.4]{cfs}).
Then, possibly after extending the particle space (see~\cite[Remark~1.2.2]{cfs}),
we can decompose the wave evaluation operator~$\Psi$ as
\beq \label{Psicompose}
\Psi = \Psi^\text{vac} + \Delta \Psi \:,
\eeq
where~$\Psi^\text{vac}$ is the wave evaluation operator of the completely filled Dirac sea
(see~\cite[\S1.1.4]{cfs} and the operator~$\Psi(x) = e^\varepsilon_x$ in~\cite[\S1.2.4]{cfs}),
and~$\Delta \Psi$ describes the holes.
The fact that the number of anti-particles is finite implies that the operator~$\Delta \Psi$
is trivial on the orthogonal complement of a finite-dimensional subspace of~$\H$, which we denote
by~$K$,
\beq \label{DelPsifinite}
\Delta \Psi \,u = 0 \qquad \text{for all~$u \in K^\perp \subset \H$}\:.
\eeq
We now choose~$(f_\tau)_{\tau \in \R}$ as the time translations, i.e.
\[ f_\tau \::\: \scrM \rightarrow \scrM \:,\: f_\tau(t,x_1,x_2,x_3) = (t+\tau, x_1, x_2, x_3) \]
(for the identification of~$\scrM$ with~$M:= \supp \rho$ see~\cite[Section~1.2]{cfs}).
Since the Lebesgue measure~$d^4x$ is translation invariant, it clearly is invariant
under the action of~$f_\tau$.
Constructing the universal measure as the push-forward (see~\cite[\S1.2.1]{cfs}), it follows immediately
that~$f_\tau$ is a symmetry of the universal measure.

Since~$\Psi^\text{vac}$ is composed of plane-wave solutions of the Dirac equation,
on which the time translation operator acts by multiplication with a phase,
the operator~$f_\tau$ can be represented by a unitary transformation in~$\H$.
More precisely, choosing the operator~$\scrU_\tau$ as the multiplication operator in momentum
space~$\hat{\scrU}(k) = e^{i k^0 \tau}$, we have the relation
\beq \label{Psivacsymm}
(\Psi^\text{vac} \,u)\big(f_\tau(x) \big) = \big( \Psi^\text{vac} \,\scrU^{-1}_\tau u \big)(x) \qquad 
\text{for all~$x \in M, u \in \H$} \:.
\eeq
Using~\eqref{Psicompose} and~\eqref{Psivacsymm} in~\eqref{Edef}, we conclude that
\[ E_\tau(u,x) := (\Delta \Psi u)\big(f_\tau(x) \big) - \big( \Delta \Psi \scrU^{-1}_\tau u \big)(x) \qquad 
\text{for all~$x \in M, u \in \H$} \:. \]
The assumption~\eqref{DelPsifinite} implies that~$(f_\tau)_{\tau \in \R}$ is indeed
a Killing symmetry with finite-dimensional support (see Definition~\ref{defkilling}).

In order to simplify the setting, we note that a unitary transformation~$\scrU_\tau$
was already used in Section~\ref{secexcurrent} to obtain corresponding conserved currents
(see~\eqref{Utaudef} and Theorem~\ref{thmcurrent}).
This means that the first variations of~$\scrU_\tau$ on the finite-dimensional subspace~$K$
give rise to a linear combination of the corresponding conserved currents.
With this in mind, we may in what follows assume that~$\scrU_\tau$ is trivial on~$K$,
\beq \label{Utrivial}
\scrU_\tau|_K = \1_K \:.
\eeq
Modifying~$\scrU_\tau$ in this way corresponds to going over to a new conservation law,
which is obtained from the original conservation law by subtracting a
linear combination of electromagnetic currents.

For the computations, it is most convenient to work again with the kernel of the fermionic projector.
Using~\eqref{Psicompose}, we decompose it as
\[ P(x,y) = P^\text{vac}(x,y) + \Delta P(x,y) \:, \]
where
\begin{align*}
P^\text{vac}(x,y) &= -\Psi^\text{vac}(x) \Psi^\text{vac}(y)^* \\
\Delta P(x,y) &= -\Psi^\text{vac}(x) \big(\Delta \Psi\big)(y)^* -\big(\Delta \Psi\big)(x) \Psi^\text{vac}(y)^*
-\big(\Delta \Psi\big)(x) \big(\Delta \Psi\big)(y)^*\:.
\end{align*}
Since~$\Delta \Psi$ vanishes on the complement of the finite-dimensional subspace~$K$,
the kernel~$\Delta \Psi$ is composed of a finite number of Dirac wave functions, i.e.
\[ \Delta P(x,y) = \sum_{i,j=1}^{\na} c_{ij} \:\phi_i(x) \overline{\phi_j(y)} \]
with~$\na \in \N$ and~$\overline{c_{ij}} = c_{ji}$. Diagonalizing the Hermitian matrix~$(c_{ij})$ by
a basis transformation, we can write~$\Delta P(x,y)$ as
\[ \Delta P(x,y) = \sum_{i=1}^{\na} c_i \:\phi_i(x) \overline{\phi_i(y)} \]
with real-valued coefficients~$c_i$.
Since we only consider first oder variations of the Lagrangian, by linearity we may restrict
attention to one of the summands. Thus it suffices to consider the case
\[ \Delta P(x,y) = \psi(x) \overline{\psi(y)} \:, \]
where~$\psi$ is a negative-frequency solution of the Dirac equation.

Now the first variation of the Lagrangian can be computed similar as in~\eqref{LTrQ} to obtain
\begin{align*}
&\frac{d}{d\tau} \int_\Omega d\rho(x) \int_{M \setminus \Omega} d\rho(y) \:\Big( \L_\kappa \big( f_\tau(x), y\big) -  \L_\kappa \big( \Phi_\tau(x), y\big) \big) \Big) \Big|_{\tau=0}  \\
&= \int_\Omega d\rho(x) \int_{M \setminus \Omega} d\rho(y)
\Big(   \Tr_{S_y} \big( Q(y,x)\, \delta_{v(x)} P(x,y) \big) + \Tr_{S_x} \big( Q(x,y)\, \delta_{v(x)} P(y,x) \big)  \Big),
\end{align*}
where
\[ \delta_{v(x)} P(x,y) := \frac{d}{d\tau} \Big( P \big( f_\tau(x) ,y \big)
-P \big( \Phi_\tau(x),y \big) \Big) \Big|_{\tau = 0} \:. \]
Since~$P^\text{vac}(x,y)$ has the Killing symmetry~\eqref{Psivacsymm},
the variation of~$P(x,y)$ simplifies to
\begin{align*}
\delta_{v(x)} P(x,y)&= \frac{d}{d\tau} \Big( \Delta P \big( f_\tau(x) ,y \big)
- \Delta P \big( \Phi_\tau(x),y \big) \Big) \Big|_{\tau = 0} \\
&= \frac{d}{d\tau} \Delta P \big( f_\tau(x) ,y \big) \big|_{\tau = 0}
= \frac{d}{d\tau} \Big( \psi \big( f_\tau(x) \big) \overline{\psi(y)} \Big) \Big|_{\tau = 0} =: (\partial_t \psi)(x) \,
\overline{\psi(y)} \:,
\end{align*}
where in the last line we used~\eqref{Utrivial}.
Using these relations in~\eqref{KillingConserve}, we obtain the conservation law
\begin{align*}
0 &= \int_\Omega d\rho(x) \int_{M \setminus \Omega} d\rho(y) 
\Big(   \Tr_{S_y} \big( Q(y,x)\, (\partial_t \psi)(x)\, \overline{\psi(y)} \big) + \Tr_{S_x} \big( Q(x,y)\, \psi(y) 
\,\overline{(\partial_t \psi)(x)}  \big)  \\
&\qquad \qquad -\Tr_{S_y} \big( Q(y,x)\,  \psi(x) \, \overline{(\partial_t \psi)(y)}
 \big) - \Tr_{S_x} \big( Q(x,y)\, (\partial_t \psi)(y)\, \overline{\psi(x)}  \big)  \Big) \\
&= 2\,\re \int_\Omega d\rho(x) \int_{M \setminus \Omega} d\rho(y)
\Big( \Sl \psi (y) | Q(y,x) (\partial_t \psi)(x) \Sr 
- \Sl \psi(x) | Q(x,y) (\partial_t \psi)(y) \Sr \Big) \:.
\end{align*}

Next, we consider the limiting case where~$\Omega$ exhausts the region between two
Cauchy surfaces~$t=t_0$ and~$t=t_1$ (see Figure~\ref{fignoether2}).
We thus obtain a conserved current~$J$ which for example at time~$t=0$
is given by
\[ J = \frac{1}{2}\: \re\,  \int_{t\leq 0} d^4x \int_{t>0} d^4y \:
\Big( \Sl \psi (y) | Q(y,x) (\partial_t \psi)(x) \Sr 
- \Sl \psi(x) | Q(x,y) (\partial_t \psi)(y) \Sr \Big) \:. \]
This equation is similar to~\eqref{inttask} and can be analyzed in exactly the same manner.
Indeed, regularizing the Heaviside functions and applying Plancherel, we
again obtain~\eqref{aneqb1} and~\eqref{aneqb2}, with the only difference
that an additional factor~$\omega_{\beta, \vec{k}}$ appears.
Thus, in analogy to~\eqref{Jsum2} and~\eqref{USymm19} we obtain
\[ J = -\sum_{\beta=1}^3 \frac{1}{2} \, c_\beta    \int \frac{d^3k}{(2 \pi)^3} \,\omega_{\beta,\vec k}^2  \:
\Sl \chi_\beta(\vec k) |  \chi_\beta(\vec k ) \Sr \:. \]
Applying~\eqref{rel2} and using again Plancherel gives the result.
This concludes the proof of Theorem~\ref{thmEMcons}.

\section{Example: Symmetries of the Universal Measure} \label{secexrho}
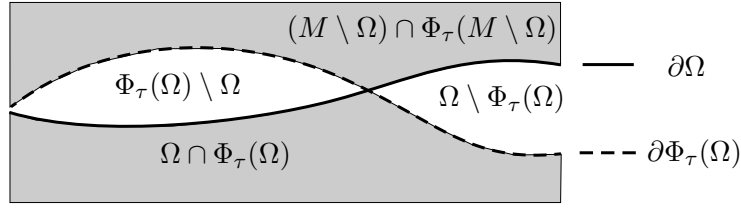
\begin{figure}
\psscalebox{1.0 1.0} 
{
\begin{pspicture}(0,-1.3321118)(9.645718,1.3321118)
\definecolor{colour0}{rgb}{0.8,0.8,0.8}
\pspolygon[linecolor=black, linewidth=0.002, fillstyle=solid,fillcolor=colour0](7.3338523,0.49955472)(6.982104,0.540414)(6.46561,0.5530239)(6.0426135,0.5034688)(5.615636,0.40687802)(5.3149023,0.32343453)(4.7772465,0.16571002)(4.124508,0.4539213)(3.5894282,0.6157628)(2.9047132,0.7188196)(1.9799689,0.7152769)(1.0838828,0.51089174)(0.40659958,0.2000194)(0.014606438,-0.059563804)(0.01385916,1.3311102)(7.3338523,1.3311102)
\pspolygon[linecolor=black, linewidth=0.002, fillstyle=solid,fillcolor=colour0](0.014606438,-0.13511193)(0.35684568,-0.23256378)(0.69938534,-0.26354155)(1.0905135,-0.30791193)(1.5389885,-0.32130453)(2.1811993,-0.3126823)(2.7668514,-0.2552749)(3.4497936,-0.16217859)(4.2947426,0.015554737)(4.786874,0.15783621)(5.7142606,-0.3485638)(6.16883,-0.5659416)(6.7202015,-0.69884527)(7.3427415,-0.68933415)(7.3346066,-1.3311119)(0.014606438,-1.3311119)
\psbezier[linecolor=black, linewidth=0.04](0.005717549,-0.13111193)(1.2346064,-0.4977786)(3.2279398,-0.25555637)(4.2857175,0.019999182)(5.3434954,0.29555473)(6.0857177,0.7111103)(7.339051,0.4999992)
\rput[bl](2.0101619,-0.9066675){$\Omega \cap \Phi_\tau(\Omega)$}
\psbezier[linecolor=black, linewidth=0.04, linestyle=dashed, dash=0.17638889cm 0.10583334cm](0.02793977,-0.057949536)(1.3812732,0.96743506)(3.2146065,0.8825633)(4.46794,0.30888808)(5.721273,-0.26478714)(6.1257176,-0.8108555)(7.339051,-0.682223)
\rput[bl](3.6946065,0.77777696){$(M \setminus \Omega) \cap \Phi_\tau(M \setminus \Omega)$}
\rput[bl](1.4057175,0.01777696){$\Phi_\tau(\Omega) \setminus \Omega$}
\rput[bl](5.716829,-0.14666748){$\Omega \setminus \Phi_\tau(\Omega)$}
\psline[linecolor=black, linewidth=0.04](7.6279397,0.50444365)(8.30794,0.50444365)
\psline[linecolor=black, linewidth=0.04, linestyle=dashed, dash=0.17638889cm 0.10583334cm](7.5879397,-0.6688897)(8.334606,-0.6688897)
\rput[bl](8.485718,-0.87555635){$\partial \Phi_\tau(\Omega)$}
\rput[bl](8.756828,0.34666586){$\partial \Omega$}
\end{pspicture}
}
\caption{The surface layer integral corresponding to a symmetry of the universal measure.}
\label{fignoether3}
\end{figure}

In this section we consider the conserved surface layer integrals corresponding to
symmetries of the universal measure (see Theorem~\ref{thmsymmum}
and Corollary~\ref{corsymmum}). We now explain why, under the assumption
that~$\Phi_\tau$ is a bijection, these conserved surface layer integrals
can be expressed merely in terms of the volumes of the sets~$\Omega \setminus \Phi_\tau(\Omega)$
and~$\Phi_\tau(\Omega) \setminus \Omega$ (see Figure~\ref{fignoether3}).
Our argument shows in particular that in the 
limiting case of Figure~\ref{fignoether2} when
the boundary of~$\Omega$ consists of two hypersurfaces, the conserved
surface layer integrals do not give rise to any interesting conservation laws.
Therefore, although the conservation laws
of Theorem~\ref{thmsymmum} and Corollary~\ref{corsymmum} give non-trivial information
on the structure of a minimizing universal measure of a causal fermion system,
they do not correspond to any conservation laws in Minkowski space.

The following argument applies for example to the situation considered in Section~\ref{seccurcor}
that~$M$ can be identified with Minkowski space, and~$\Omega$ is the past of a Cauchy surface.
But the argument applies in a much more general setting. In particular, we do not need to
assume that~$\Omega$ is compact. We first rewrite the surface layer integral in~\eqref{conserve5} as
\begin{align*}
\int_\Omega& d\rho(x) \int_{M \setminus \Omega} d\rho(y) \:\Big( \L_\kappa\big( \Phi_\tau(x), y\big) -
\L_\kappa\big( x, \Phi_\tau(y) \big) \Big) \\
&= \int_\Omega d\rho(x) \int_{M \setminus \Omega} d\rho(y) \:\Big( \L_\kappa\big( \Phi_\tau(x), y\big) -
\L_\kappa(x, y) \Big) \\
&\quad + \int_\Omega d\rho(x) \int_{M \setminus \Omega} d\rho(y) \:\Big( \L_\kappa(x, y) -
\L_\kappa\big( x, \Phi_\tau(y) \big) \Big) \\
&= \bigg( \int_{\Phi_\tau(\Omega)} - \int_\Omega \bigg) \:d\rho(x) \int_{M \setminus \Omega} d\rho(y) \: \L_\kappa(x, y) \\
&\quad + \int_\Omega d\rho(x) \:\bigg( \int_{M \setminus \Omega} - \int_{\Phi_\tau(M \setminus \Omega)} \bigg)\:
d\rho(y) \:\L_\kappa(x, y) \:.
\end{align*}
Assuming that~$\Phi_\tau$ is as bijection, we can write the obtained differences of integrals as
\begin{align*}
\bigg( \int_{\Phi_\tau(\Omega)} - \int_\Omega \bigg) \;\cdots =
\bigg( \int_{\Phi_\tau(\Omega) \setminus \Omega} - \int_{\Omega \setminus \Phi_\tau(\Omega)} \bigg) \;\cdots \\
\bigg( \int_{M \setminus \Omega} - \int_{\Phi_\tau(M \setminus \Omega)} \bigg) \;\cdots =
\bigg( \int_{\Omega \setminus \Phi_\tau(\Omega)} - \int_{\Phi_\tau(\Omega) \setminus \Omega} \bigg) \;\cdots
\end{align*}
(see Figure~\ref{fignoether3}).
We thus obtain
\begin{align*}
\int_\Omega& d\rho(x) \int_{M \setminus \Omega} d\rho(y) \:\Big( \L_\kappa\big( \Phi_\tau(x), y\big) -
\L_\kappa\big( x, \Phi_\tau(y) \big) \Big) \\
&= \bigg( \int_{\Phi_\tau(\Omega) \setminus \Omega} - \int_{\Omega \setminus \Phi_\tau(\Omega)} \bigg)
\:d\rho(x) \int_M d\rho(y) \: \L_\kappa(x, y) \\
&= \bigg( \int_{\Phi_\tau(\Omega) \setminus \Omega} - \int_{\Omega \setminus \Phi_\tau(\Omega)} \bigg)
\:\ell(x)\: d\rho(x) \:,
\end{align*}
where in the last step we used~\eqref{elldef}. In view of~\eqref{ELgen}, the obtained integrand
is constant. Therefore, the surface layer integral can indeed be expressed in terms of the volume of the
sets~$\Phi_\tau(\Omega) \setminus \Omega$ and~$\Omega \setminus \Phi_\tau(\Omega)$.
In particular, the surface layer integral does not capture any interesting dynamical information
of the causal fermion system.

\section{Outlook: Conservation Laws in Quantum Space-Times} \label{secoutlook}
We again point out that the conservation laws of
Theorem~\ref{thmcurrent} and Theorem~\ref{ThmKillingCons} hold for causal
fermion systems without taking the continuum limit.
In particular, these conservation laws also hold for regularized Dirac sea
configurations if one analyzes the EL equations corresponding to the causal
action principle without taking the limit~$\varepsilon \searrow 0$.
We now give an outlook on potential implications of these conservation
laws in such ``quantum space-times''.

\begin{Remark} {\bf{(conservation laws and microscopic mixing)}} \label{remmicro} {\em{
In Theorem~\ref{thmcurrentmink} and Theorem~\ref{thmEMcons} we restricted attention
to negative-frequency solutions of the Dirac equation. On a technical level, this was
necessary because the operator~$\hat{Q}$ is only well-defined inside the {\em{lower}}
mass cone, whereas it diverges outside the lower mass cone (see Definition~\ref{def611}
and~\cite[Section~5.6]{PFP} or~\cite{reg}). In non-technical terms, this means that
introducing Dirac particles into the causal fermion system makes the causal action infinitely large.
But, as explained in detail in~\cite[Section~3]{qft}, the action becomes again finite if
one introduces a so-called {\em{microscopic mixing of the wave functions}}.
In other words, minimizing the causal action gives rise to the mechanism of microscopic mixing
(for more details see~\cite[\S1.5.3]{cfs}).
Microscopic mixing is also important for getting the connection to entanglement
and second-quantized bosonic fields (see~\cite{entangle, qft, qed}).

If microscopic mixing is present, the conservation laws of
Theorem~\ref{thmcurrent} and Theorem~\ref{ThmKillingCons} again give rise to
conserved surface layer integrals. However, evaluating these surface layer integrals
in Minkowski space is more involved because a homogenization procedure over the
microstructure must be performed (in the spirit of~\cite[Section~5.1]{qft}).
Since these constructions are rather involved, we cannot give them here.
However, even without entering the detailed constructions, the following argument
shows that the conservation laws should apply to the particle states as well:

In an interacting system, a solution of the Dirac equation which at some initial time
has negative frequency may at a later time have positive frequency
(as in the usual pair production process).
The conservation law of Theorem~\ref{thmcurrent} implies that the surface layer
integral at the later time coincides with that at the initial time.
Using current conservation of the Dirac dynamics, we conclude that the
surface layer at the later time again coincides with the surface integral
of the Dirac current.
Using arguments of this type, one sees that, no matter
what the microscopic structure of space-time is, the conservation laws
of Theorem~\ref{thmcurrent} and Theorem~\ref{ThmKillingCons} should
apply similarly to positive-frequency solutions of the Dirac equation.
\QEDrem }} \end{Remark}

\begin{Remark} {\bf{(conservation laws and collapse)}} \label{remcollapse} {\em{
As explained in~\cite[Section~3]{dice2010} and~\cite[Section~7]{dice2014}, there are strong indications that
an analysis beyond the continuum limit leads to nonlinear effects
which allow for the description of the collapse of the wave function in the quantum mechanical
measurement process. In this context, the results of this paper have the following
implications: 

Suppose that the wave function undergoes a collapse at some time~$t_c$.
Then at this time, the system cannot be described by the continuum limit.
However, it is a reasonable assumption that the continuum limit should still be a good
description at some earlier time~$t_0<t_c$ and some later time~$t_1>t_c$.
Similar as explained in Remark~\ref{reminteract},
in this situation the conservation law of Theorem~\ref{thmcurrent}
states that the current integrals at times~$t_0$ and~$t_1$ coincide.
In other words, the conservation law of Theorem~\ref{thmcurrent} implies that the
collapse mechanism necessarily preserves the normalization of the wave function.
Thus, in contrast to some continuous dynamical localization models
(see for example~\cite{ghirardi2, pearle0, bassi}), in our approach it is not necessary
to rescale the wave function so as to arrange its proper normalization.

More generally, Theorem~\ref{thmsymmgis2} implies that all
other conservation laws obtained in the continuum limit
are also respected by the collapse. In particular, Theorem~\ref{ThmKillingCons}
gives conservation of energy-momentum in the collapse process.
\QEDrem }} \end{Remark}

\Thanks {{\em{Acknowledgments:}}
J.K.\ gratefully acknowledges support by the ``Studienstiftung des deutschen Volkes.''

\newpage

\begin{thebibliography}{10}

\bibitem{barutbook}
A.O. Barut, \emph{Electrodynamics and {C}lassical {T}heory of {F}ields \&
  {P}articles}, Dover Publications, Inc., New York, 1980, Corrected reprint of
  the 1964 original.

\bibitem{bassi}
A.~Bassi, K.~Lochan, S.~Satin, T.P. Singh, and H.~Ulbricht, \emph{Models of
  wave-function collapse, underlying theories, and experimental tests}, Rev.
  Mod. Phys. \textbf{85} (2013), 471--527.

\bibitem{lagrange}
Y.~Bernard and F.~Finster, \emph{On the structure of minimizers of causal
  variational principles in the non-compact and equivariant settings},
  arXiv:1205.0403 [math-ph], Adv. Calc. Var. \textbf{7} (2014), no.~1, 27--57.

\bibitem{davies}
B.~Davies, \emph{Integral {T}ransforms and their {A}pplications}, third ed.,
  Texts in Applied Mathematics, vol.~41, Springer-Verlag, New York, 2002.

\bibitem{cfs}
F.~Finster, \emph{The {C}ontinuum {L}imit of {C}ausal {F}ermion {S}ystems},
  book based on the preprints arXiv:0908.1542 [math-ph], arXiv:1211.3351
  [math-ph] and arXiv:1409.2568 [math-ph], in preparation.

\bibitem{PFP}
\bysame, \emph{The {P}rinciple of the {F}ermionic {P}rojector}, hep-th/0001048,
  hep-th/0202059, hep-th/0210121, AMS/IP Studies in Advanced Mathematics,
  vol.~35, American Mathematical Society, Providence, RI, 2006.

\bibitem{reg}
\bysame, \emph{On the regularized fermionic projector of the vacuum},
  arXiv:math-ph/0612003, J. Math. Phys. \textbf{49} (2008), no.~3, 032304, 60.

\bibitem{continuum}
\bysame, \emph{Causal variational principles on measure spaces},
  arXiv:0811.2666 [math-ph], J. Reine Angew. Math. \textbf{646} (2010),
  141--194.

\bibitem{entangle}
\bysame, \emph{Entanglement and second quantization in the framework of the
  fermionic projector}, arXiv:0911.0076 [math-ph], J. Phys. A: Math. Theor.
  \textbf{43} (2010), 395302.

\bibitem{dice2010}
\bysame, \emph{The fermionic projector, entanglement, and the collapse of the
  wave function}, arXiv:1011.2162 [quant-ph], J. Phys.: Conf. Ser. \textbf{306}
  (2011), 012024.

\bibitem{qft}
\bysame, \emph{Perturbative quantum field theory in the framework of the
  fermionic projector}, arXiv:1310.4121 [math-ph], J. Math. Phys. \textbf{55}
  (2014), no.~4, 042301.

\bibitem{cfsrev}
\bysame, \emph{Causal fermion systems -- an overview}, arXiv:1505.05075
  [math-ph], to appear in {Q}uantum {M}athematical {P}hysics: A {B}ridge
  between {M}athematics and {P}hysics (F.~Finster, J.~Kleiner, C.~Röken, and
  J.~Tolksdorf, eds.), Birkh\"auser Verlag, Basel, 2016.

\bibitem{lqg}
F.~Finster and A.~Grotz, \emph{A {L}orentzian quantum geometry},
  arXiv:1107.2026 [math-ph], Adv. Theor. Math. Phys. \textbf{16} (2012), no.~4,
  1197--1290.

\bibitem{vacstab}
F.~Finster and S.~Hoch, \emph{An action principle for the masses of {D}irac
  particles}, arXiv:0712.0678 [math-ph], Adv. Theor. Math. Phys. \textbf{13}
  (2009), no.~6, 1653--1711.

\bibitem{jet}
F.~Finster and J.~Kleiner, \emph{The jet bundle dynamics of causal fermion
  systems}, in preparation.

\bibitem{dice2014}
\bysame, \emph{Causal fermion systems as a candidate for a unified physical
  theory}, arXiv:1502.03587 [math-ph], J. Phys.: Conf. Ser. \textbf{626}
  (2015), 012020.

\bibitem{support}
F.~Finster and D.~Schiefeneder, \emph{On the support of minimizers of causal
  variational principles}, arXiv:1012.1589 [math-ph], Arch. Ration. Mech. Anal.
  \textbf{210} (2013), no.~2, 321--364.

\bibitem{qed}
F.~Finster and J.~Tolksdorf, \emph{A microscopic derivation of quantum
  electrodynamics}, in preparation.

\bibitem{norm}
\bysame, \emph{Perturbative description of the fermionic projector:
  Normalization, causality and {F}urry's theorem}, arXiv:1401.4353 [math-ph],
  J. Math. Phys. \textbf{55} (2014), no.~5, 052301.

\bibitem{forger+roemer}
M.~Forger and H.~R{\"o}mer, \emph{Currents and the energy-momentum tensor in
  classical field theory: a fresh look at an old problem},
  arXiv:hep-th/0307199, Ann. Physics \textbf{309} (2004), no.~2, 306--389.

\bibitem{ghirardi2}
G.C. Ghirardi, P.~Pearle, and A.~Rimini, \emph{Stochastic processes in
  {H}ilbert space: a consistent formulation of quantum mechanics}, Foundations
  of quantum mechanics in the light of new technology ({T}okyo, 1989), Phys.
  Soc. Japan, Tokyo, 1990, pp.~181--189.

\bibitem{goldstein}
H.~Goldstein, \emph{Classical {M}echanics}, second ed., Addison-Wesley
  Publishing Co., Reading, Mass., 1980.

\bibitem{halmosmt}
P.R. Halmos, \emph{Measure {T}heory}, Springer, New York, 1974.

\bibitem{hawking+ellis}
S.W. Hawking and G.F.R. Ellis, \emph{The {L}arge {S}cale {S}tructure of
  {S}pace-{T}ime}, Cambridge University Press, London, 1973.

\bibitem{landau2}
L.D. Landau and E.M. Lifshitz, \emph{The {C}lassical {T}heory of {F}ields},
  Revised second edition. Course of Theoretical Physics, Vol. 2. Translated
  from the Russian by Morton Hamermesh, Pergamon Press, Oxford, 1962.

\bibitem{noetheroriginal}
E.~Noether, \emph{Invariante {V}ariationsprobleme}, Nachr. d. {K}\"onig.
  {G}esellsch. d. {W}iss. {M}ath-phys. {K}lasse, Berlin (1918), 235--257.

\bibitem{pearle0}
P.~Pearle, \emph{Combining stochastic dynamical state-vector reduction with
  spontaneous localization}, Phys. Rev. A \textbf{39} (1989), no.~5,
  2277--2289.

\bibitem{straumann}
N.~Straumann, \emph{General {R}elativity}, Texts and Monographs in Physics,
  Springer-Verlag, Berlin, 2004.

\end{thebibliography}
\providecommand{\bysame}{\leavevmode\hbox to3em{\hrulefill}\thinspace}
\providecommand{\MR}{\relax\ifhmode\unskip\space\fi MR }
\providecommand{\MRhref}[2]{%
  \href{http://www.ams.org/mathscinet-getitem?mr=#1}{#2}
}
\providecommand{\href}[2]{#2}

\end{document}